\documentclass[x11names]{article}

\pdfoutput=1

\usepackage{arxiv}

\usepackage[bookmarks=false]{hyperref}

\usepackage{pdfpages}

\usepackage{palatino}

\usepackage[utf8]{inputenc} 
\usepackage[T1]{fontenc}    
\usepackage{url}            
\usepackage{booktabs}       
\usepackage{amsfonts}       
\usepackage{nicefrac}       
\usepackage{microtype}      
\usepackage{lipsum}		
\usepackage{graphicx}
\usepackage[round]{natbib}
\usepackage{doi}

\usepackage{figsize}
\usepackage{amssymb}
\usepackage{amsmath}

\usepackage{dcolumn}

\usepackage{tikz}
\usetikzlibrary{matrix,calc}
\usetikzlibrary{positioning}
\usepackage{pdflscape}

\usepackage{nameref}

\usepackage{cleveref}

\usepackage{booktabs}
\usepackage{longtable}

\usetikzlibrary{decorations.pathreplacing}

\definecolor{myLightGray}{RGB}{191,191,191}
\definecolor{myGray}{RGB}{160,160,160}
\definecolor{myDarkGray}{RGB}{144,144,144}
\definecolor{myDarkRed}{RGB}{167,114,115}
\definecolor{myBlue}{RGB}{84,175,215}
\definecolor{myGreen}{RGB}{0,255,71}

\usepackage{mwe}

\usepackage{erewhon}
\usepackage{lipsum}
\usepackage{lettrine}
\usepackage{GoudyIn}

\usepackage{xcolor}

\LettrineTextFont{\itshape}
\usepackage{blindtext}
\usepackage{tocbibind}

\hypersetup{
colorlinks=true, 
linktoc=all,     
 linkcolor=magenta, citecolor= blue 
}

\usepackage{lineno}

\usepackage{graphicx}
\usepackage{floatrow}

\title{On spatial variation in the detectability and density of social media user protest supporters} 

\usepackage[noblocks]{authblk}

\author[1]{Víctor H. Masías\thanks{Corresponding author: vmasias@fen.uchile.cl.}}
\author[2]{Fernando Crespo}
\author[3]{Pilar Navarro R.}
\author[1]{Razan Masood}
\author[4]{Nicole C. Krämer}
\author[1]{H. Ulrich Hoppe}

\affil[1]{Department of Computer Science and Applied Cognitive Science, University of Duisburg--Essen, Germany}
\affil[2]{Facultad Tecnológica, Universidad de Santiago de Chile, Estación Central, Santiago, Chile}
\affil[3]{E.T.S. of Computer and Telecommunication Engineering, University of Granada, Spain}
\affil[4]{Social Psychology: Media and Communication, University of Duisburg--Essen, Germany}

\date{}
\renewcommand{\headeright}{}

\begin{document}
\maketitle
\begin{abstract}
Although much has been published regarding street protests on social media, few works have attempted to characterize social media users' spatial behavior in such events. The research reported here uses spatial capture-recapture methods to determine the influence of the built environment, physical proximity to protest location, and collective posting rhythm on variations in users' spatial detectability and density during a protest in Mexico City. The best-obtained model, together with explaining the spatial density of users, shows that there is high variability in the detectability of social media user protest supporters and that the collective posting rhythm and the day of observation are significant explanatory factors.  The implication is that studies of collective spatial behavior would benefit by focussing on users' activity centres and their urban environment, rather than their physical proximity to the protest location, the latter being unable to adequately explain spatial variations in users' detectability and density during the protest event.    
\end{abstract}
	
\section{Introduction}

\lettrine{P}{rotest events}{}  are a social phenomenon that contributes to processes of change in all political systems, whether democratic, non-democratic, or some hybrid of the two. In the last years, massive demonstration events held in the USA have generated renewed international attention from the scientific community on the phenomenon of street protest \citep{fisher2019science}.  They typically take place in major urban centres and are inspired by a wide range of different motives. Recently, it has been found that there is a causal relationship between the moralization processes that occur in social media and the behavior of individuals in an offline environment  \citep{mooijman2018moralization}. Little is known, however, about the spatial behavior of social media users who support street protests and how the urban environment of a city influences it. This paper focuses on examining in greater detail the spatial behavior of social media user protest supporters at the city level.

Previous works have explored protest behavior in online environments and its correlations with the spatial dimension of offline protest \citep{chen2012you,Traag_2017,mooijman2018moralization}. Part of the contemporary research which studies protests in natural settings argues that the physical proximity to the protest location is an important explanatory of the spatial behavior of supporters during protest events. For example, it has been found that the place of residence, that is, if the social media user is frequently observed in the center on the periphery of the city is correlated with the level of support expressed in social media to a protest event \citep{chen2012you,Barber__2015}. Also, studies based on socio-physical models have proposed that physical distance to the protest location acts as an impedance to attendance \citep{Traag_2017}. Other studies have proposed that the relationship between environmental characteristics and the distribution of individuals across space is more complex and dynamic, prompting researchers to investigate the relationship using simulation models \citep{davies2013mathematical,lemos2017protestlab,pires2017modeling,bacaksizlar2019understanding}, whose results are, therefore, of unknown ecological validity.  

In this context, it is relatively unexplored whether existing elements of the urban environment influence the spatial behavior of social media user protest supporters. For this reason, we propose to conduct an observational interdisciplinary study that applies an ecological approach, known as spatial capture-recapture, to test a series of alternative hypotheses that theoretically have a better explanatory capacity in the context of social media research. Within this scenario, our research question is whether existing structures in the urban environment, such as street or subway networks, or whether the rate of social media posting in a given geographic area have significant ability to explain the detectability (i.e. the probability of detecting a user in a given place at a given occasion) and spatial density (i.e. the numbers of users divided by a spatial area) of the social media users supporting the protest. Thus, the objectives of this study can be summarized as follows: 

\begin{itemize} 
\item Determine the factors that explain the variation in spatial detectability of social media user protest supporters.

\item Determine the factors that explain the variation in spatial density of social media user protest supporters.

\item Determine whether physical proximity to a protest location contributes to explaining the detectability and density of social media user protest supporters.

\item Compare different models and evaluate whether social media post rhythms in a given geographical region, transport network structures, the proximity of the users to the protest location, neighborhood-level socio-demographic variables, and the day of observation, contributes to explaining the detectability and density of social media user protest supporters.

\end{itemize} 

To determine these subjects, we will use multiple types of information that are generally difficult to use together under the same methodological framework. Our focus will be on protests at the city level to explain relationships between the spatial dimension and social media user behavior. More specifically, we examine how protests events are reflected in the spatial behavior of social media users who support a protest.

\section{Conceptual-Analytical Framework} 

Although the relationship between environmental elements and social media users' spatial behavior has only recently attracted the attention of researchers, earlier work on the field of animal ecology had already thrown light on the impact of urbanization on species richness and diversity \citep{mckinney2008effects}. Chronobiology research on cattle wandering the streets of India's cities has found that their activity patterns (i.e. lying down, standing, walking, foraging) are correlated with environmental factors \citep{sahu2019spatiotemporal}. In the case of human beings, there is a complex interaction between land use and human activities that take place in the urban environment. Other studies have proposed that socio-ecological systems maintain reciprocal interactions between biophysical and socioeconomic structures \citep{arnaiz2018identifying}. In short, those studies suggest that there is a significant relationship between the environment and organisms' spatial behavior. 

In this context, research in the existing literature suggests that the spatial behavior of individuals is influenced by factors related to both the individuals themselves and the geographic area they live in. In particular, numerous works have documented the way basic processes such as daily physical activity, cognitive performance, and locomotion are regulated by circadian rhythms \citep{valdez2019homeostatic}. In evolutionary terms, it has been suggested that chrono-physiological processes control and facilitate the organization of spatial and temporal behavior in living beings. As an example, animal behavior studies indicate that wallabies follow circadian patterns in their search for food and shelter to avoid dangers in their environment \citep{fischer2019circadian}. 

Recently, in the field of social media research, it has been shown that daily posting rates of users can be used as a proxy for the endogenous circadian clock of individuals. For example, the daily use of social media has been used to examine how the rate of collective posting varies seasonally and geographically, and also to further demonstrate that social events and pressures disrupt patterns of social media activity,  as occurs in the so-called ``Twitter social jet lag'' phenomenon \citep{leypunskiy2018geographically}. In another research more consistent with chronobiology studies, empirical evidence has been reported on the existence of the phenomenon of social synchronization in Facebook Messenger \citep{Diwan_2020}. Generally speaking, research suggests that the behavior of posting social media content can be used as a proxy measure for the endogenous circadian clock of users of social networking sites \citep{murnane2015social,kumar2020habitual}. In light of the above, it seems plausible to consider that the daily posting rates of social media content in a given geographical region can be a relevant factor that can help explain variations in the spatial behavior of social media user protest supporters.

A parallel line of research has focused on investigating how the transport network configurations influence individuals' spatial behavior. This research has found, for example, that street centrality is positively correlated with different types of land use \citep{Rui_2014} and can be a good predictor of pedestrian flow \citep{Bielik_2018}. Other authors have shown that street network configurations can explain serious outdoor violence \citep{Summers_2016}. These findings imply that street centrality may affect individual spatial behavior, which in turn means it is reasonable to expect that the centrality of such spatial structures would also have the potential to explain variations, across a geographical region, of the spatial behavior of social media user protest supporters.

Finally, neighborhood-level socio-demographic characteristics have been cited as having an impact on the spatial distribution of individuals over time and space. For example, a comparative study found that a set of geographical characteristics of the built environment consistently accounted for the greater part of the variability in population distribution in low- and medium-income countries \citep{Nieves_2017}. Other studies have determined that a range of urban factors explain the density of mobile telephone activity within cities \citep{De_Nadai_2016,Liu_2020}. In general terms, it seems plausible that both neighborhood-level socio-demographic characteristics and the built environment have the potential to explain variations of the spatial behavior of social media user protest supporters.

With the foregoing in mind and to enhance our understanding of the spatial behavior of social media user protest supporters, we designed an observational study based on the spatial capture-recapture approach \citep{royle2013spatial} (see `Spatial capture-recapture analysis' in the Methods section). More specifically, our approach attempted to determine whether and to what extent the physical proximity of the social media users (hereafter ``SMUs'' or simply ``users'') to the protest location, socio-demographic characteristics, street network configurations, and the daily rhythms of social media posting in a given geographical region, and the day of observation,  contribute in accounting for the variations in the detectability and density of SMUs at the city level. The event used as a real case study is the 40th annual Mexico City LGBT pride parade held on June 23, 2018. The organizers of the event proposed a march that has a distance of $\approx$4km. The main meeting place was planned to be in the \textit{Angel of Independence}\footnote{This is a frequent starting point for various demonstrations and not only for this case study (see, \url{https://en.wikipedia.org/wiki/Angel_of_Independence}).} at 10:00 a.m. From there, different groups marched to reach the so-called \textit{Zócalo}\footnote{This city square has an area of 57,600m$^2$ and is located in the downtown area of Mexico City (see, \url{https://en.wikipedia.org/wiki/Zócalo}).}, where a concert was held during the afternoon, for concluding the whole event. This case is of particular interest because the parade is one of the largest mass gatherings held in the Mexican capital \citep{bosia2020oxford}. And unlike other types of collective behavior such as riots, the events organized by the Mexican LGBT movement are primarily of a collaborative, non-violent nature \citep{beer2018extending}.\footnote{This social movement has achieved in Mexico several successes in the political sphere such as the legalization of same-sex civil unions and LGBT adoptions, demonstrating its effectiveness in bringing about changes in human rights law. These changes have not only increased the acceptance of LGBT persons in Mexican society as a whole but have also improved their perspectives for embarking upon and developing careers in the sciences \citep{walker2014equality}.}

\begin{figure}
   \centering
\hspace*{-10pt}
\begin{tikzpicture}[every node/.style={anchor=north east,fill=white,minimum width=.5cm,minimum height=5mm}]
\matrix (mA) [draw,matrix of math nodes]
{
(1) & (0) & (1) & (0) \\
(0) & (1) & (0) & (1) \\
(1) & (0) & (1) & (0) \\
(0) & (1) & (0) & (1) \\
};
\matrix (mB) [draw,matrix of math nodes] at ($(mA.south west)+(1.5,0.7)$)
{
(0) & (1) & (0) & (1) \\
(1) & (0) & (0) & (0) \\
(0) & (1) & (0) & (1) \\
(1) & (0) & (1) & (1) \\
};

\matrix (mC) [draw,matrix of math nodes] at ($(mB.south west)+(1.5,0.7)$)
{
(1) & (0) & (1) & (0) \\
(0) & (1) & |[draw=myBlue]|(1) & (1) \\
(1) & (0) & (1) & (0) \\
(0) & (1) & (0) & (0) \\
};
\draw[dashed](mA.north east)--(mC.north east);
\draw[dashed](mA.north west)-- node[sloped,above] {Occasions ($k$ $=$ 1,..., $K$)} (mC.north west);
\draw[dashed](mA.south east)--(mC.south east);
\node [above left] at (mC.north) {Traps ($j$ $=$ 1,..., $J$)};
\node [left] at (mC.west) {Users ($i$ $=$ 1,..., $I$)};

\draw (mC-2-3) to[out=660,in=200] +(6cm,-3cm) node[left] {A social media user was captured posting something today, here and now };

\end{tikzpicture}

\begin{tikzpicture}[%
    every node/.style={
        font=\scriptsize,
        text height=1ex,
        text depth=.25ex,
    },
]
\draw[->] (0,0) -- (13,0);

\foreach \x in {0,4,...,13}{
    \draw (\x cm,4pt) -- (\x cm,0pt);
}

\node[anchor=north] at (2,0) {$t-1$};

\node[anchor=north] at (6,0) {$t$};

\node[anchor=north] at (10,0) {$t+1$};

\node[anchor=north] at (12.5,0) {\texttt{sessions} };

\fill[myLightGray] (0,-0.4) rectangle (4,-0.55);
\fill[myBlue] (4,-0.4) rectangle (8,-0.55);
\fill[myGreen] (6,-0.4) rectangle (6,-0.55);
\fill[myLightGray] (8,-0.4) rectangle (12,-0.55);

\draw[decorate,decoration={brace,amplitude=7pt}] (4,-0.8) -- (8,-0.8)
    node[anchor=north,midway,below=4pt] {Protest event day};
\end{tikzpicture}

\caption{Indexing social media users in a 3--dimensional array.}
\label{EDA}
\end{figure}
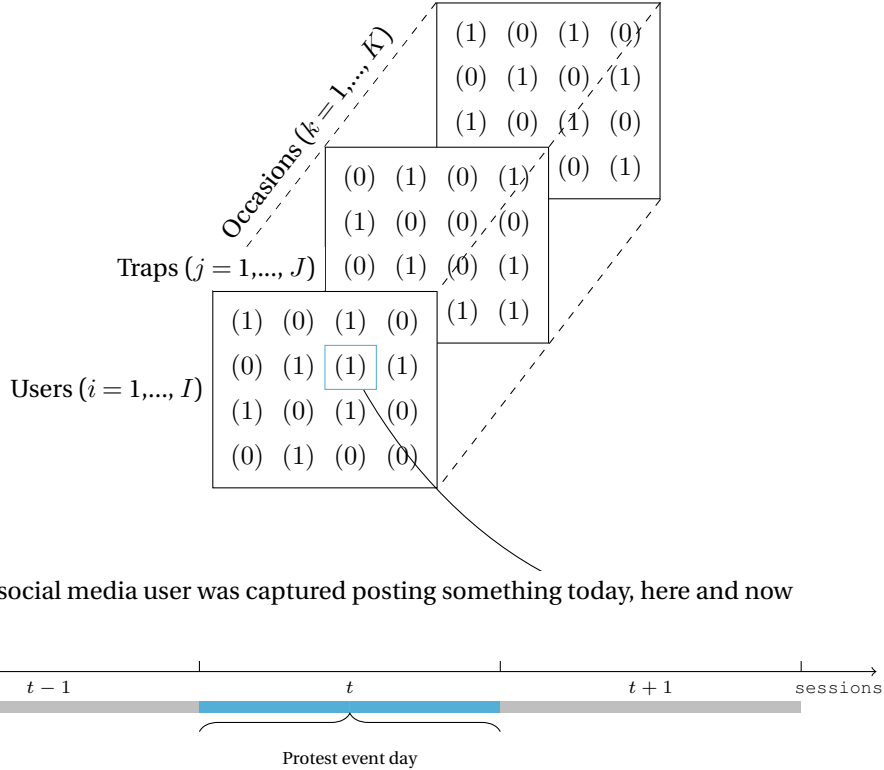

We used geotagged tweets to study the spatial behavior of users who supported this protest event. The geotagged social media data for the study were collected through the Twitter API, and the tweets were filtered based on geolocation information and constrained to Mexico City. To operationalize which SMUs supported the march, we coded 10,000 tweets by hand and used them to train a logistic regression classifier that then identified supportive tweets in our whole geotagged social media data sample (see `Identification of users supporting the protest event' in the Methods section). To organize the data for the analysis of SMU's spatial behavior during the march, we imposed a spatial grid over Mexico City in which each grid cell, hexagonal and covering 1.18km$^2$, represented a trap where SMUs could be captured or recaptured in time and space (see Figure \ref{EDA}). For each such SMU capture or recapture, we indexed in a 3--dimensional array who ($i$), where ($j$), and when ($k$). Thus, $y_{i, j, k}$ $=$ 1 denoted an individual user who was captured in a given cell on a given occasion, and $y_{i, j, k}$ $=$ 0 indicated that the individual was not captured in that cell on that occasion. To identify each SMU ($i$), we use the unique identifier already assigned to each user by the Twitter API (the \texttt{User ID} field) and which was anonymized to maintain the user's privacy. A unique identifier was also assigned to each grid cell ($j$) and each of the 24 hours in a day ($k$). 

Besides, three sets of proxies were used to serve as metrics of the urban environment factors and the SMU's attributes hypothesized to influence the spatial detectability and density of SMU protest supporters. The first set of proxy measures corresponds to the concept of the physical proximity of the SMUs to the protest location (see `Physical proximity to the protest location' in the Methods section). The second set consisted of a single proxy, which was the Twitter post rate of SMUs in the Mexican capital, which was measured using the \texttt{Tweetogram} metric \citep{leypunskiy2018geographically} that expresses the normalized activity rate of users over 24 hours (see `Measuring the tweeting rhythm' in the Methods section). The third set measured was the centrality of the Mexico City street and metro network. The network data were obtained from OpenStreetMap using the automated approach developed by \citet{boeing2017osmnx,Boeing_2020}. The metro network was based on the official information published by the Mexican metro system. The specific proxies used for quantifying street intersection centrality were the degree, betweenness, and closeness measures for weighted networks \citep{opsahl2010node} (see ``Measuring street and metro networks centralities' in the Methods section). The fourth set represented socio-demographic and socio-economic characteristics based on an anonymized data sample from the 2010 Mexican census of 58,064 blocks in Mexico City. One of these proxies measured population density while the other was a relative wealth index based on an aggregate measure of household assets, both calculated at the city block level (see `Demographic and economic indices' in the Methods section). Our database consisted of $i$ $=$ 1,216 SMUs who supported the march and who were captured by $j$ $=$ 158 traps over 3 consecutive observation days (i.e. \texttt{sessions}) encompassing the day of the march and one day on either side.

The testing and evaluation of SMU's detectability and density variations in the sample were done using spatial capture-recapture methods \citep{royle2013spatial}, which is an approach developed in the field of population and landscape ecology. This approach has definite advantages for identifying patterns in hard-to-find samples given that, even if the latter are small and incomplete, it can provide estimates of their detectability and density over a given geographical area. The method makes joint estimates of a pair of models, one for SMU's detectability and the other for SMU's spatial density (see Table \ref{Parameters}). The detections decay as a function of the distance between the trap and the detected SMU's activity centre $d(s)$, the latter considered as a latent variable \citep{royle2018unifying}. Thus, for the detection model, we use the so-called half--normal detection function. It has two parameters: baseline detection probability $p_{0}$, which determines the maximum detection probability at the SMU's activity centre, and spatial decay $\sigma$, which controls how rapidly detection probabilities decrease with distance from it  (see `Spatial capture-recapture analysis' in the Methods section). 

\begin{table*}
\centering
\begin{tabular}{@{}lll@{}}
\toprule
Model & Parameter & Formula   \\ \midrule
Baseline encounter model &  $p_{0}$ &  logit$\left(p_{0}\right)=\alpha_{0}+\sum_{k} \alpha_{k} Cov_{k}$  \\
 &  & \\
Spatial decay model & $ \sigma$ & $\log (\sigma)=\gamma_{0}+\sum_{k} \gamma_{k} Cov_{k}$  \\
 &  & \\
Spatial point process model  &   $d(s)$  & $\log (d(s))$=$\beta_{0}+\sum_{k} \beta_{k} Cov_{k}$ \\ 
\bottomrule
\end{tabular}
\caption{Parameters of the SMU's  detectability and density variation models. $Cov_k$ are the different covariates affecting the parameters and $\alpha_k$, $\gamma_k$, $\beta_k$ are the covariate coefficients.}
\label{Parameters}
\end{table*}

These parameters are theoretically significant for understanding SMU's spatial behavior during protest events given that $d(s)$ allows us to estimate the variation if any, in SMU's spatial density, $p_{0}$ allows us to estimate the maximum detectability of SMUs during a protest event, and with $\sigma$ we can estimate any variation in the use of a space before, during and after the protest day. The importance of estimating these parameters lies in the fact that previous studies, especially those in sociology, have tended to focus on estimating the local number (i.e. at the protest location) of persons attending a demonstration \citep{rotman2020using} without also considering the changing detectability, use of space, and density. This is relevant because it is usually not possible to make a direct estimation of user spatial density with geotagged social media data, and especially because the sampling mechanism of social media APIs is generally unknown. Also, the user's location can only be observed when posting, so his or her activity centre is not directly observable, and the estimation of space use needs to be addressed differently.  Also, given the understanding of SMU's spatial behavior developed here, observed changes in density, detectability, or spatial decay on the protest day could be seen as an expression of the collective engagement. By the same token, a lack of significant change in the group's spatial parameters could be said to indicate that there had been no change in its spatial behavior and level of engagement during the protest event. The approach proposed here is thus very powerful in its ability to identify behavior patterns using observational samples of SMUs and spatial covariates.

Several candidate model configurations were used to test what factors explain the detectability and density of the observed SMUs.  First, a null model was created, where the density of social media user, the baseline probability of detection, and the scale parameter is kept constant over the city. Second, a series of alternative models were created. For the baseline encounter model ($p_{0}$), the measures tested were demographic and economic indices of the neighborhood, the centrality of the street and metro network, the \texttt{Tweetogram} measure (which measures the rate of posting in the city for each hour), the day of observation (i.e. \texttt{session}), and the physical proximity between the endpoint of the protest march and the trap where the SMU was captured or recaptured posting social media content. For the spatial decay model ($\sigma$) we tested whether observation day (i.e. \texttt{session}) contributed to explaining variability in the use of space. Additionally, for the density model ($d$($s$)), we include the demographic and economic indices of the neighborhood, the centrality of the street and metro network, the day of observation (i.e. session), and the physical proximity between the endpoint of the march and SMU's possible activity centres (i.e. represented by a regular grid of points). 

Maximum likelihood was used to jointly estimate the model parameters and evaluate which candidate model best fit the data. Specifically, a likelihood analysis of various models were conducted using the R package \textsc{oSCR}  \citep{sutherland2019oscr}, a type of generalized linear mixed model. As can be seen, this methodological design is quite flexible given that maximum likelihood allows us to compare multiple competing models and explanatory spatial and temporal factors.

\section{Methods}

\subsection{Spatial capture-recapture analysis} The following account of the spatial capture-recapture method is based on the work of \citet{sutherland2019oscr} which was adapted to our research context. In conceptual terms, an SMU is assumed to have an activity centre ($s_{i}$ $=$ [$s_{i,X}$,$s_{i,Y}$]) that can be regarded as a spatial coordinate. However, an SMU's location can only be known if it is captured in a trap, so $s_{i}$ is considered to be a latent rather than an observable variable. SMUs are also presumed to live within a geographic area represented by a state space ($S$). In our methodological setting, the null model specifies that each SMU's activity centre is distributed uniformly in space:

\begin{equation}
s_i \sim Uniform (S)
\label{UNIFORME}
\end{equation}

To determine the activity centre values we use maximum likelihood estimation with spatial capture-recapture (SCR) models. This approach is based on the marginal likelihood that the unknown variable \textbf{s} is removed by averaging (or marginalizing) over the possible values of \textbf{s}. Thus, we begin by identifying the \textit{conditional-on-s} likelihood. The user encounter model $y_{ijk}$ for individual $i$ at trap $j$ on occasion $k$, conditional upon $s_i$, is

\begin{equation}
y_{ijk} | s_i \thicksim Bernoulli( p ( x , s_i;  \theta ) )
\end{equation}

where
\begin{equation}
p (x ,s) = p_0\; exp (- \| x-s \|^ {2}/( 2\sigma^{2}) )
\end{equation}

is the detectability or probability of detection at trap \textbf{x}, which depends on \textbf{s} and $\theta = (p_0,\sigma)$ (see Table \ref{Parameters} for more details). The joint distribution of the data for individual $i$ is the product of $J \times K$ such terms (i.e., contributions from each of $J$ traps and $K$ occasions):

\begin{equation}
[ y_i | s_i , \theta ] =\prod_{j=1}^{J} \prod_{k=1}^{K}  Bernoulli (p ( x_j , s_i ; \theta ))
\end{equation}

This assumes that an encounter of individual $i$ in each trap and on each occasion is independent of encounters in every other trap and occasion, conditional upon \textbf{$s_i$}.

To compute the marginal likelihood, consider a regular grid of $G$ points denoted $s_u$ that form cells of equal area and are indexed by $u $=$ 1, 2, ..., G$. In our design, we use a grid of 479 points. The marginal probability mass function (pmf) of $y_i$ is then approximated by

\begin{equation}
[y_i|\theta]=\frac{1}{G}\sum_{u=1}^{G} [y_i|s_u,\theta] 
\end{equation}

The joint likelihood for the data from $n$ observed individuals, assuming independence of encounters among and between individuals, is the product of $n$ such terms and a contribution of the $n_0$ $=$ $N-n$ uncaptured individuals. Each of these all-zero encounter histories (i.e., the capture histories of individuals not encountered) will have the same marginal pmf contribution in the likelihood given above, here denoted by $\pi_0$. Of the $n$ observed individuals, there are $N \choose n$ = $\frac{N!}{n!n_0!}$ ways to choose a sample of size $n$. A combinatorial term is therefore required, the joint likelihood then being given by

\begin{equation}
\mathcal{L}\left(\theta,n_{0}\mid\mathbf{y}\right)=\frac{N!}{n!n_0!}\left\{\prod_{i=1}^{n}\left[\mathbf{y}_i\mid\theta\right]\right\}\pi_{0}^{n_0}
\end{equation}

Our data are from distinct and more or less independent populations, referred to as \texttt{sessions}, that correspond to the three days in June considered in the study. The multi-session model integrates data from these different \texttt{sessions} . If $N_g$ is the population size of group $g$, there are $(\frac{N_{g} !}{n_{g} !\left(N_{g}-n_{g}\right) !})$ ways to choose a sample. The multi-session model assumes that

%

$$N_g \thicksim Poisson(\lambda_g)  $$

where the $N_g$ are mutually independent random variables. We obtain the marginal likelihood as a function of $\lambda_g$ by independently marginalizing over this distribution of $N_g$ for the data from each group:

\begin{equation}
\mathcal{L}(\lambda_g,\theta)=\sum_{g=1}^\infty \mathcal{L}(\theta,n_{0_g}|\mathbf{y})Poisson(N_g;\lambda_g)
\end{equation}

Explicit models can be formulated for $\lambda_g$:
$$log(\lambda_g)=\beta_0+\beta_1Cov_g$$

Spatial covariates can be obtained by raster data or by inverse distance-weighted interpolation, as in this study, which also contributes to maintaining offline the privacy of the studied geographical areas. For further details on the assumptions of spatial capture-recapture modelling, see \citet{royle2013spatial,sutherland2019oscr}.

\subsection{Identification of users supporting the protest event}
\label{coding}

The identification of SMUs who supported the protest was accomplished in two stages. The first stage was to manually label the data using qualitative coding in a small sample of social media data. The second stage utilized a machine learning approach to search in a larger sample for protest-related tweets.

\subsubsection{Qualitative coding of data}
\label{Qualitative}

The data coding procedure consisted of five steps that were applied to the text of each tweet, as follows. Step 1: Determine whether it includes hashtags relating to the march. Step 2: Determine whether it contains emoticons related to this social movement. Step 3: Determine whether it has any other content supporting the protest (e.g., photos, videos, maps, or any other media content in the URL). Step 4: Read the entire text to determine whether it suggests support for the protest. Step 5: Code each tweet based on the results of the previous steps. This last step was carried out by two raters. If a tweet contained references to the protest, it was coded \texttt{YES}, otherwise, it was coded \texttt{NO}.\footnote{Additional details of the coding can be found in \cite{Masias_2019}.} The outcome of this procedure was that, of the 10,000 coded tweets, 796 users supporting the march. The value of Cohen's kappa coefficient for the two raters' coding results ($\kappa$$=$0,84) indicated a high level of agreement between them. This initial codification was used for the next section.

\subsubsection{Machine learning classification}
\label{Labelling}

Once the 10,000-item database was codified, the classification of the entire sample using machine learning could be performed. The tweet texts to be classified were first lemmatized and cleaned by removing stop words, punctuation, URLs, and mentions. Hashtags were kept and emoticons were replaced with code words given that they played an important role in distinguishing related tweets. Logistic regression was used to classify the tweets and a grid search was used to refine the model parameters. We then applied a term-frequency transformer and performed a feature selection using the chi-square test to choose the most significant 500 terms. Because the dataset is unbalanced, we also employed combined under-sampling and over-sampling techniques using SMOTE+Tomek as suggested in \citet{santos2018cross}. The performance of the classifiers (see Table \ref{PERFORMANCE}) was evaluated by 5-fold cross-validation on the test dataset measured in term of ROC-AUC, as well as the Macro-F1 score (M-F1) and the Matthews correlation coefficient (MCC), the last two often used to measure performance with unbalanced databases. The obtained model was then used to make predictions regarding the non-coded data in the sample. Logistic regression obtained 2,605 protest-related tweets and 114,848 unrelated ones. After filtering the database, the number of users identified as supporters on each of the three days covered by the data (i.e. 22, 23, and 24 of June) was 526, 846, and 539 respectively. Therefore, the use of machine learning made it possible to identify additional users supporting the protest march. 

\begin{table}[ht]
\centering
\begin{tabular}{@{}llll@{}}
\toprule
Classifier & M-F1 & ROC-AUC  & MCC \\ \midrule
Logistic regression & 0.95 & 0.98 & 0.91 \\

\bottomrule
\end{tabular}
\caption{Summary of classification performance.}
\label{PERFORMANCE}
\end{table}

\subsection{Measuring the tweeting rhythm}

To measure SMU's  tweeting activity in the city we used the  measure (A(t)) \citep{leypunskiy2018geographically}, defined as: 

\begin{equation}
A(t)=\frac{1}{N} \sum_{i} \frac{f_{i}(t)}{\sum_{\tau} f_{i}(\tau)}
\end{equation}

where $N$ is the number of observed users normalized to a unit weight, and $f_{i}$ is the number of tweets posted by user $i$ during a time bin $t$, in this case, 1 hour. The time series were then averaged across the 72 hours (i.e., the three observation days, see Figure \ref{8por16}).

\begin{figure}
\centering
\includegraphics[width=14cm]{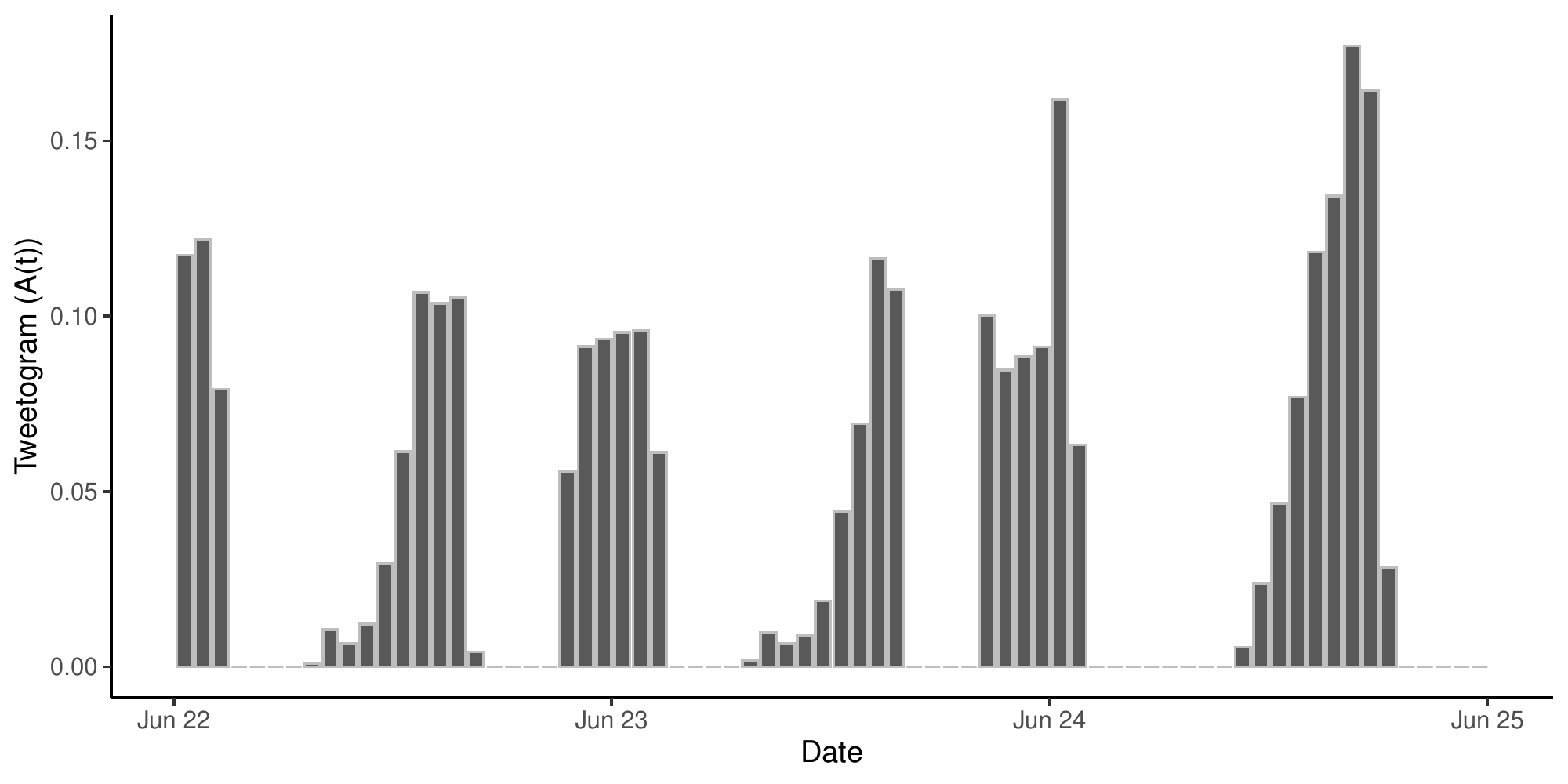} 
\caption{Normalized posting rate of users in Mexico City during three days of observation. On June 23, the day of the planned protest event, the maximum value of \texttt{Tweetogram} was reached at 3:00 p.m.}
\label{8por16}
\end{figure}

\subsection{Measuring street and metro networks centralities}
\label{centrality}

A graph of the Mexico City street network was generated that contained 112,188 nodes representing intersections and 164,586 edges representing streets. The mean and standard deviation of the edge lengths were 88.91 meters and $SD$ $=$ 128.12, respectively. To measure street centrality as an explanatory factor of SMU's density in the city, we used the centrality metrics $\alpha\omega$-weighted degree, $\alpha\omega$-weighted betweenness, and $\alpha\omega$-weighted closeness \citep{opsahl2010node}. These indices allowed us to measure the centrality of street intersections, and metro stations (i.e. nodes)  depending on the ``number'' or the ``strength'' of the links, or both. A tuning parameter $\alpha$ was incorporated to include additional information in the centrality measures. The values used for this parameter were: $\alpha$  $= $ 0, which leaves the centrality measures in their traditional form; $\alpha$  $= $ 0.5, which weights the measures by the number of edges and the latter's weights (the lengths of the streets in meters, and the distance between pairs of connected metro stations); and $\alpha$  $= $ 1, which weights the measures only by the street lengths.\footnote{Further information on how these definitions of the weighted centrality measures compare with the original, non-weighted definitions may be found in \citet{masias2017exploring}. Network data is publicly available from OpenStreetMap through \textsc{OSMnx} Package. See,   \citet{boeing2017osmnx,Boeing_2020}.}

\subsection{Demographic and economic indices}
\label{indices}

The neighborhood-level socio-demographic indicators were created with anonymized data from the 2010 Mexican census.\footnote{The anonymized census data of Mexico City was made publicly available by Diego del Valle  (\url{https://blog.diegovalle.net/2013/06/shapefiles-of-mexico-agebs-manzanas-etc.html}).} Household and dwelling attributes were used to create indices at the census block level, which were then normalized by census-block area. 

\subsubsection{Population density per census block} 

This indicator described the population density of each census block in terms of its population density per dwelling and household. The information was aggregated and normalized by census-block area. A principal components analysis (PCA) was carried out to detect independent features characterizing the population density per census block. We used the first two main components (i.e. \texttt{population density PC1} and \texttt{population density PC2}) that captured 64.3\% and 32.6\% percent of the total variation to create a feature vector that characterized the population density characteristics of each census block (n=58,064) of Mexico City.

\subsubsection{Relative wealth per census block} 

This index measured the relative wealth of a census block.  It was constructed from census data on two categories of assets. The first category was household items such as information and communications technology devices (radios, televisions, computers, landlines, and mobile telephones, etc.), refrigerators, washing machines, and motor vehicles. The second category included dwelling characteristics such as internet access, electricity, running water, toilets, and connection to the city sewage system. The asset counts were aggregated by census block and normalized by census-block area.  In the same way, a principal components analysis (PCA) was carried out to detect independent features characterizing the relative wealth per census block. The first two main components (i.e. \texttt{wealth PC1} and \texttt{wealth PC2}) that captured 95.9\% and 3.1\% percent of the total variation were used to construct a feature vector characterizing the relative wealth of each census block (n = 58,064) in Mexico City.

\subsection{Physical proximity to the protest location}
\label{DISTANCE}

Finally, we created two spatial covariates to operationalize the concept of physical proximity. The first one was the Haversine distance between the place where the social media user was captured or recapture posting social media content in a given trap and the centroid of the Zócalo. The second measure was the Haversine distance between the centroid of the Zócalo and the and SMU's initial possible activity centres (i.e. for this, we used the regular grid of 479 points defined in section `Spatial capture-recapture analysis'). As can be seen, these two distance measures will be used as covariates for the detectability model, and the density model, respectively.

\section{Results}

In the following, we report the results on capture history, model fitting and selection, and the main results obtained by studying the spatial behavior of SMU protest supporters.

\subsection{Capture history description} 

The number of SMU protest supporters who were captured on the three consecutive observation days (Friday, Saturday, and Sunday) was 526, 846, and 539. The mean maximum distance moved (MMDM) was 7110.91 m (excluding cases where the distance was zero). A visual representation of captured geotagged social media data is given in Figure \ref{8Captured}, which shows for each day where the SMUs were captured in the grid cells imposed over Mexico City. 

\begin{figure}
\includegraphics[width=0.3\textwidth]{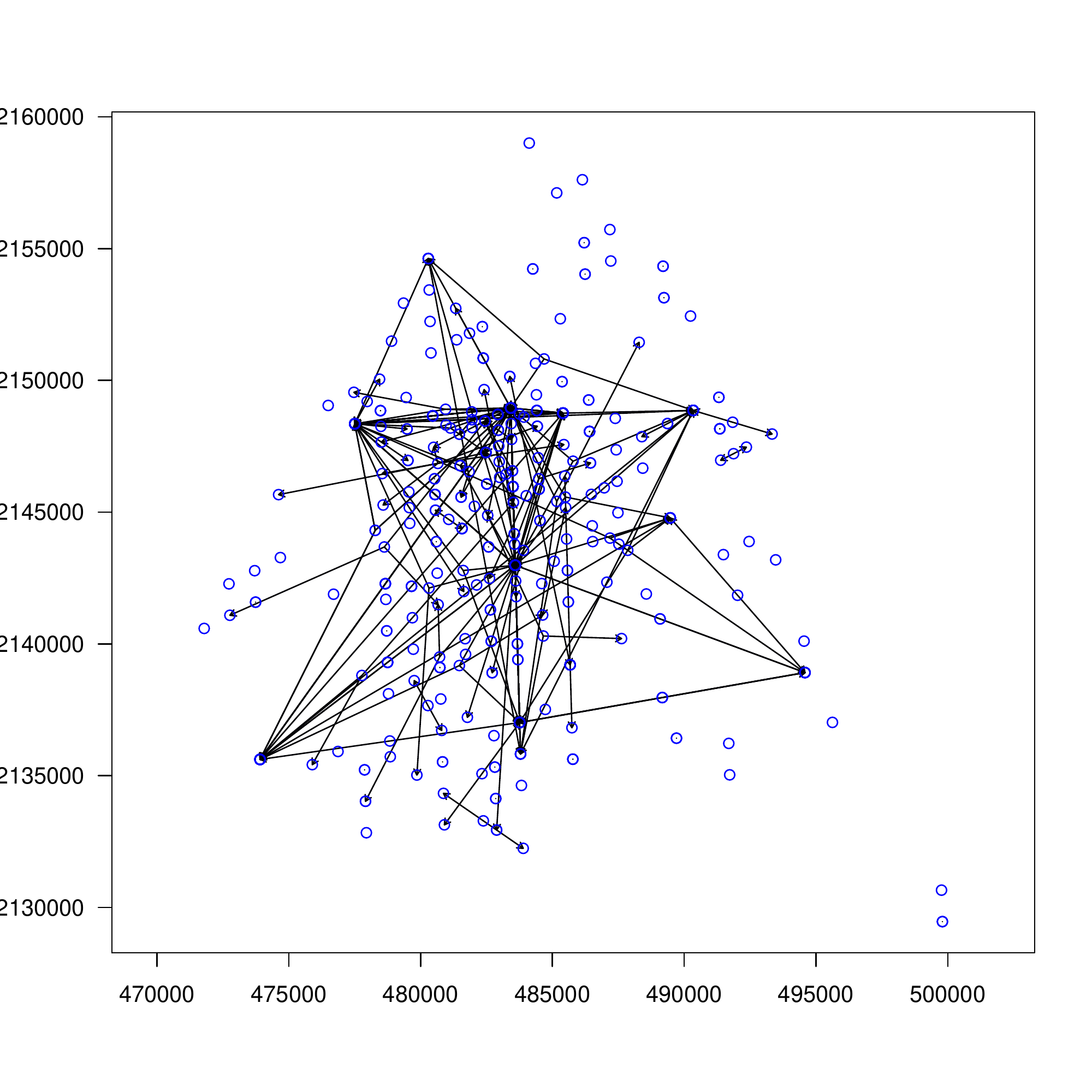}\hfil
\includegraphics[width=0.3\textwidth]{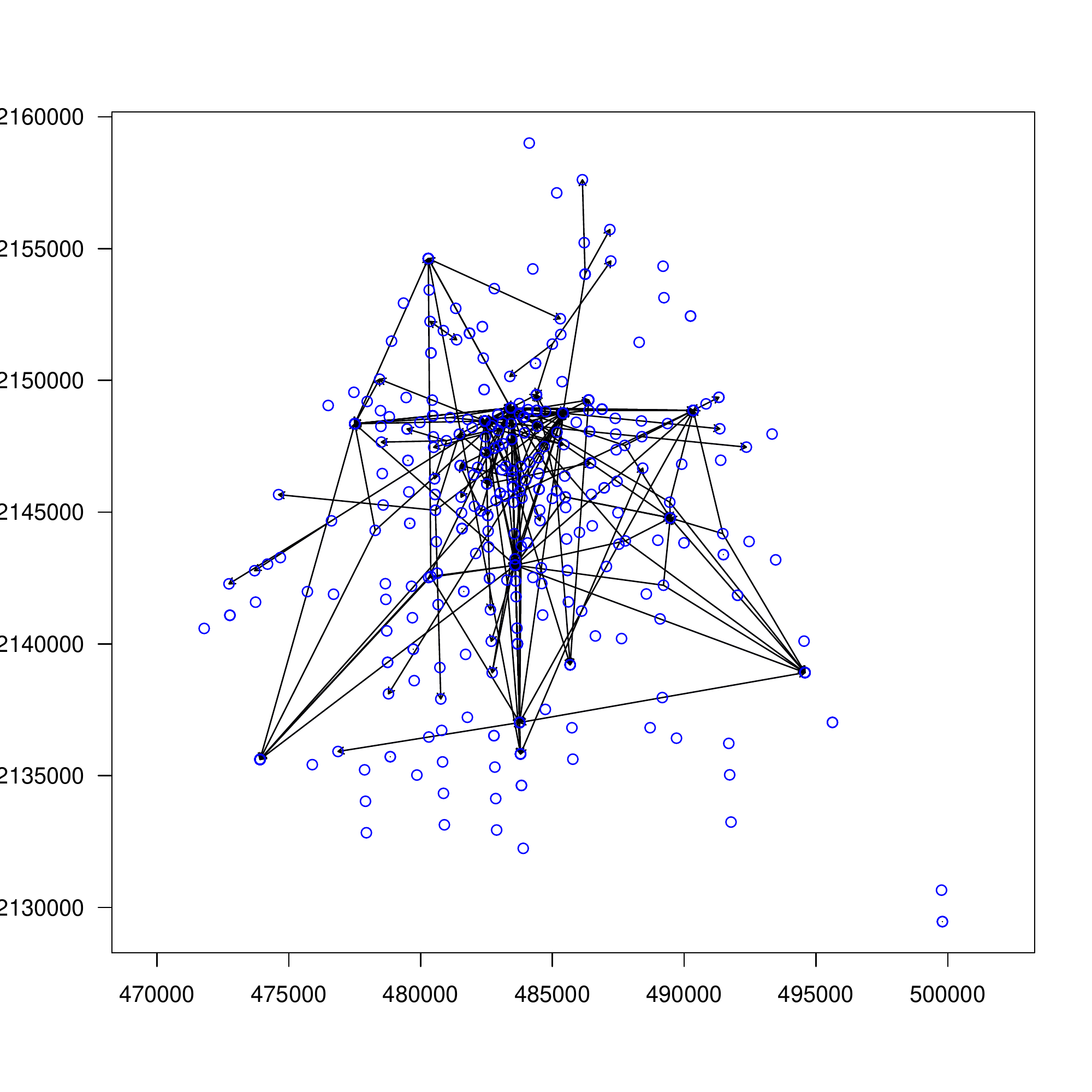}\hfil
\includegraphics[width=0.3\textwidth]{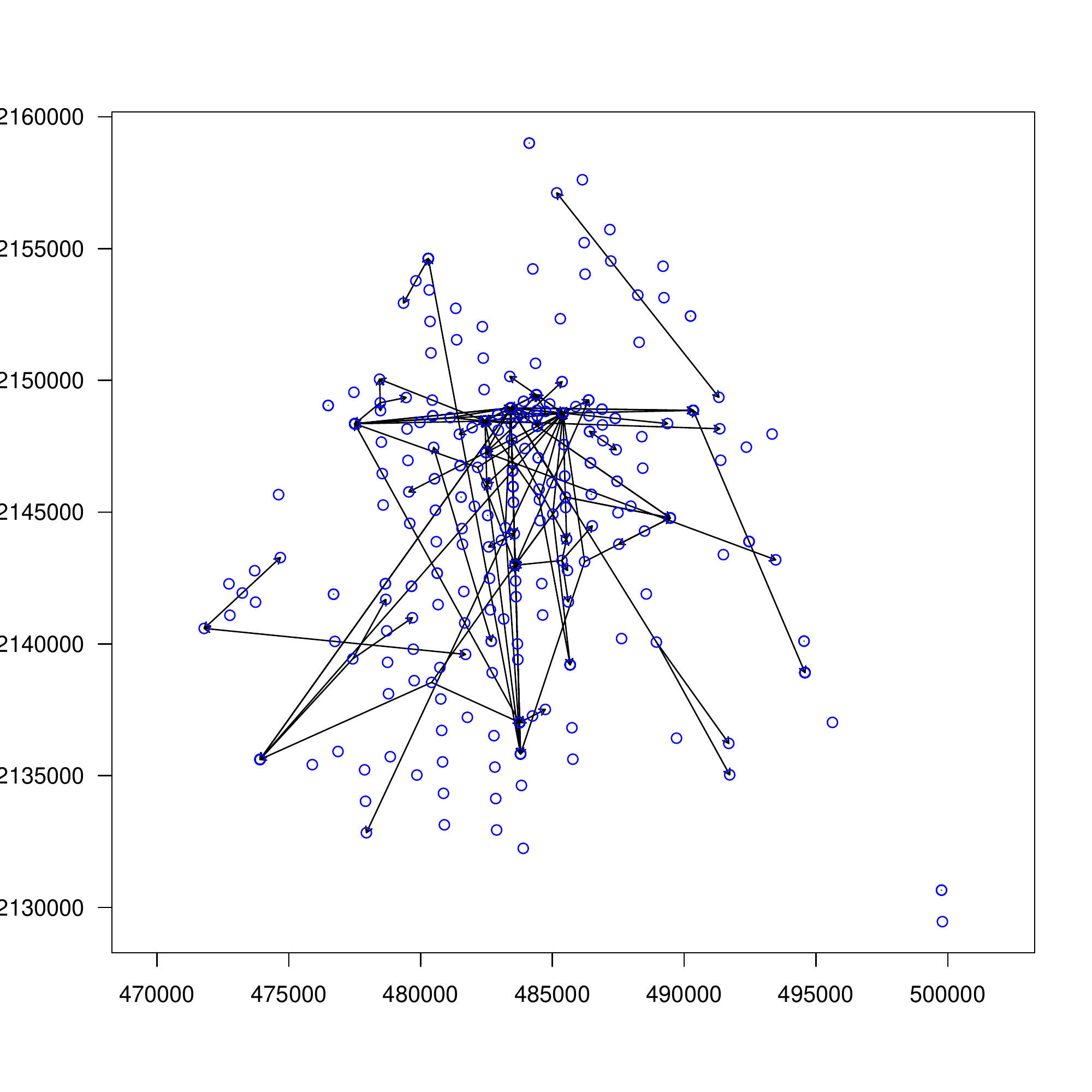}
\medskip
\hspace{0.5\baselineskip}
\makebox[0.3\textwidth]{\texttt{ session 1} }\hfil
\makebox[0.3\textwidth]{\texttt{ session 2} }\hfil
\makebox[0.3\textwidth]{\texttt{ session 3} }
\setcounter{figure}{2}
\caption{The capture history was constructed from the data collected on social media in Mexico City. Each blue circle represents the cell/trap centroid, and the directed arrows connect pairs of traps where SMUs were captured and recaptured on the indicated observation day, where  \texttt{session 1} corresponds to June 22, \texttt{session 2} corresponds to June 23, and \texttt{session 3} corresponds to June 24. The circles with a dot at the centre indicate that the user was captured only by that trap while those without a dot indicate that no user was captured on that day. }
\label{8Captured}
\end{figure}

From a qualitative point of view, it can be seen that during \texttt{session 2} the captures and recaptures are centered in the up center of the map and that it corresponds to the downtown of Mexico City where the protest was planned.

\subsection{Model fitting and selection}

The conceptual and methodological approach developed allows generating evidence--based comparative results. The statistic used to determine the best model was the Akaike Information Criterion (AIC). The AIC values were calculated for each candidate model and the differences ($\Delta$AIC) were used to rank them, the model with the lowest value being the one with the greatest explanatory power. The results for each obtained model are presented in  Table \ref{fit}. To present the results, below are the results of 8 models (out of a total of 100 configuration models tested): the best model obtained; the null model used for comparison purposes; the models based exclusively on the hypothesis of physical proximity to the protest location; a model that explains the density and detectability exclusively from the day of observation (i.e. only by \texttt{session});  and finally 3 models with the lowest performance obtained.

\begin{table}[ht!]
\centering

\resizebox{\textwidth}{!}{%
\begin{tabular}{@{}llllllllll@{}}
\toprule
Model & Density ($d(s)$ )                & Baseline detection probability ($p_{0}$)                  & Spatial decay ($\sigma$)  & logL     & K  & AIC       & $\Delta$AIC       & $\Omega$     & CumWt     \\* \midrule
                                                                
\textbf{Best }& $\sim\alpha_{0}-\omega\texttt{weighted street degree} + \alpha_{0}\omega-\texttt{weighted metro betweenness}$  & $\sim\texttt{Tweetogram} + \texttt{session}$ & $\sim\texttt{session}$ &  16956.01	& 10 & 	33932.01 & 	  0	&  9.999993e-01	 & 0.9999993 \\

Alternative   &  $\sim\texttt{session}$  & $\sim\texttt{session}$ &  $\sim\texttt{session}$ & 19044.19 &	 9 & 	38106.38 & 4174.3708	 & 0.000000e+00 &	1\\

Alternative   &  $\sim\texttt{Proximity to the protest location}$   & $\sim\texttt{1}$ & $\sim\texttt{1}$ & 19051.59	& 4	& 38111.18 & 4179.17055	& 0.000000e+00  & 1 \\

\textbf{Null} &                                                                         $\sim\texttt{1}$ & $\sim\texttt{1}$ & $\sim\texttt{1}$  & 19123.79	&  3	& 38253.58 &  4321.56967	&  0.000000e+00 & 	1  \\

Alternative &  $\sim\texttt{1}$ & $\sim\texttt{Proximity to the protest location}$ & $\sim\texttt{1}$ & 19315.61	& 4	& 38639.23 & 4707.21099	& 0.000000e+00 &	 1 \\
Alternative     & $\sim\alpha_{0.5}\omega-\texttt{street betweenness}$ & $\sim\texttt{Proximity to the protest location}$ & $\sim\texttt{session}$   & 19629.07 &	 7	& 39272.14 & 5340.12337	&  0.000000e+00	 & 1 \\
Alternative     & $\sim\alpha_{0}\omega-\texttt{street betweenness}$ & $\sim\texttt{Proximity to the protest location}$ & $\sim\texttt{session}$   & 19662.77 &	7 &	39339.53 & 5407.51729 &	0.000000e+00	& 1 \\
Alternative   &                             $\sim\texttt{population density PC2}$ & $\sim\texttt{Proximity to the protest location}$ & $\sim\texttt{session}$ & 19662.77 &	7 &	39339.53 & 5407.51942 &	0.000000e+00 &	1 \\
\bottomrule
\end{tabular}}
\caption{Summary of model selection for estimating variation in density ($d(s)$), baseline detection probability ($p_{0}$) and spatial decay ($\sigma$) of  SMU protest supporters. Indicated are the Akaike Information Criterion (AIC), the ranking based on AIC ($\Delta$AIC) and model weights ($\Omega$). Here, $\alpha_0$ refers to $\alpha=0$, $\alpha_{0.5}$ corresponds to a value of $\alpha=0.5$, and $\alpha_1$ refers to $\alpha=1$ for the weighted centrality measures.\label{fit}} 
\end{table}

Several observations can be made when comparing the models listed in Table ~\ref{fit}. First, Models using the \texttt{proximity to the protest location} performed similarly or less than the null model. Second, if we fit a model to explain the density and detectability exclusively from the day of observation (i.e. \texttt{session}), we also obtain a model comparable to the null model. Third, the three lowest-performing models have in common that they used the \texttt{proximity to the protest location} for the baseline detection probability. In the following, the best model obtained is described.

\subsection{Description of the model}The comparative analysis of different models shows that density ($d(s)$) varies with street degree centrality and metro betweenness centrality, the baseline detection probability ($p_{0}$) varies with observation day (i.e. \texttt{session}) and the \texttt{Tweetogram} variable, and spatial decay ($\sigma$) varies with observation day. The AIC of the best model obtained was 33932.01 and the model weights were $\Omega$ = 9.999993e-01. The results for the best model, summarized in Table ~\ref{SUMMARY}, indicate that all of these variables were statistically significant. 

\setcounter{table}{3}
\begin{table*}[h!]
\centering
\caption{Summary of best model results (two-tailed $p$--value).}
\begin{tabular}{@{}lllllll@{}}
\toprule
Parameters & Estimate    & SE    &   z & P(\textgreater{}$\mid$ z $\mid$)     \\ \midrule
p0.(Intercept)                     & -7.302 & 0.067 & -108.927  & 0.000 \\
p0.\texttt{session 2}                         &   0.170 & 0.074 &   2.301  & 0.021 \\
p0.\texttt{session 3}                        &  -0.597 & 0.092 &  -6.469  & 0.000 \\
t.beta.\texttt{Tweetogram}         &  20.520 & 0.385 &  53.318  & 0.000 \\
sig.(Intercept)                    &   8.068 & 0.027 & 296.672  & 0.000 \\
sig.\texttt{session 2}                      &  -0.062 & 0.036 &  -1.700  & 0.089 \\
sig.\texttt{session 3}                      &  -0.214 & 0.044 &  -4.800  & 0.000 \\
d0.(Intercept)                     &   0.907 & 0.495 &   1.832  & 0.067 \\
d.beta.$\alpha_{0}\omega$-\texttt{weighted street degree}             &  -1.052 & 0.066 & -15.871  & 0.000 \\
d.beta.$\alpha_{0}\omega$-\texttt{weighted metro betweenness}  &  0.376 & 0.071 &   5.256  & 0.000 \\
\bottomrule
\end{tabular}
\label{SUMMARY}
\end{table*}

\subsection{Spatial detectability and density variation results}

\begin{figure}
\centering
\SetFigLayout{2}{2}
  \subfigure[]{\includegraphics{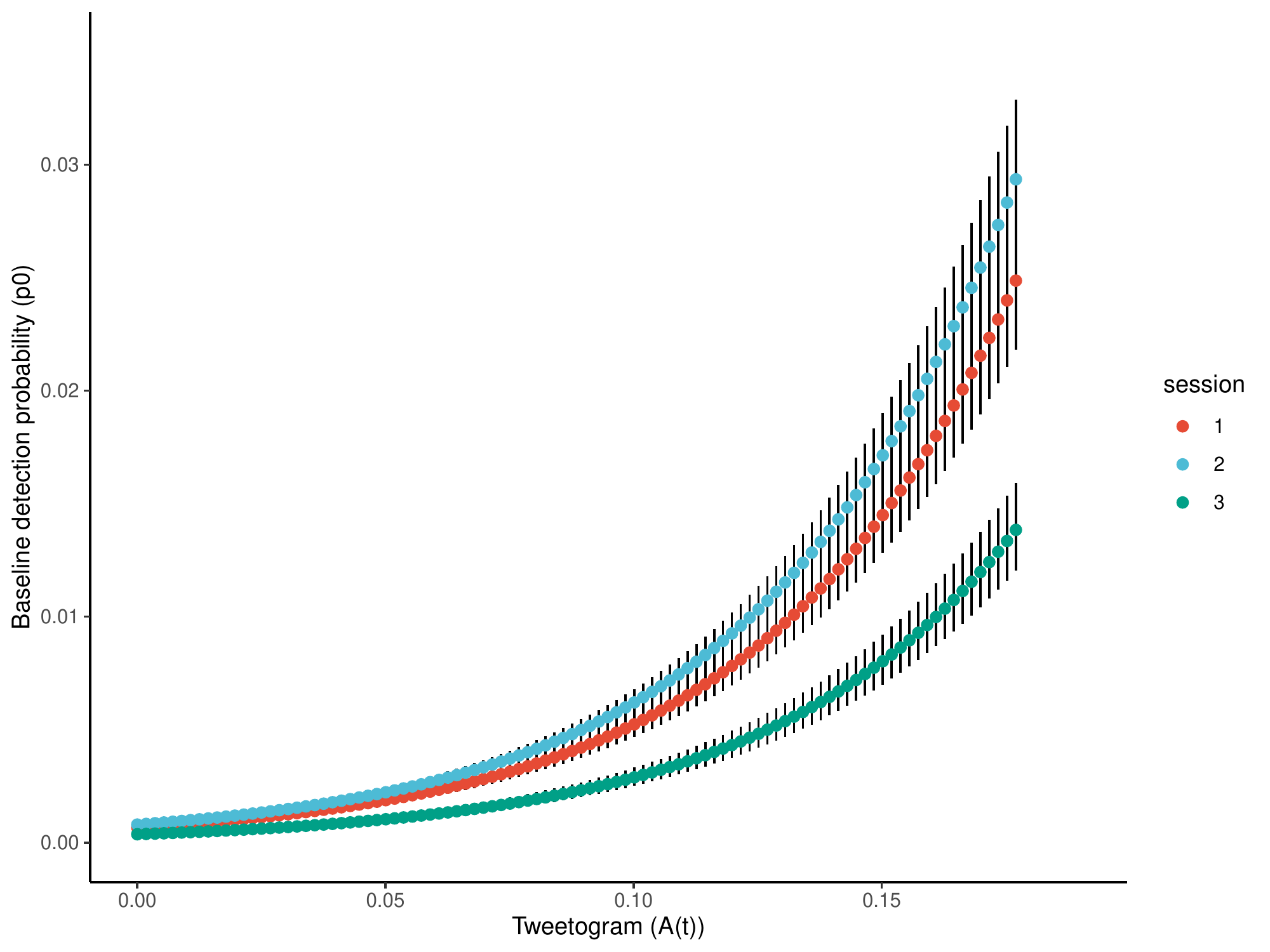}}
  \hfill
  \subfigure[]{\includegraphics{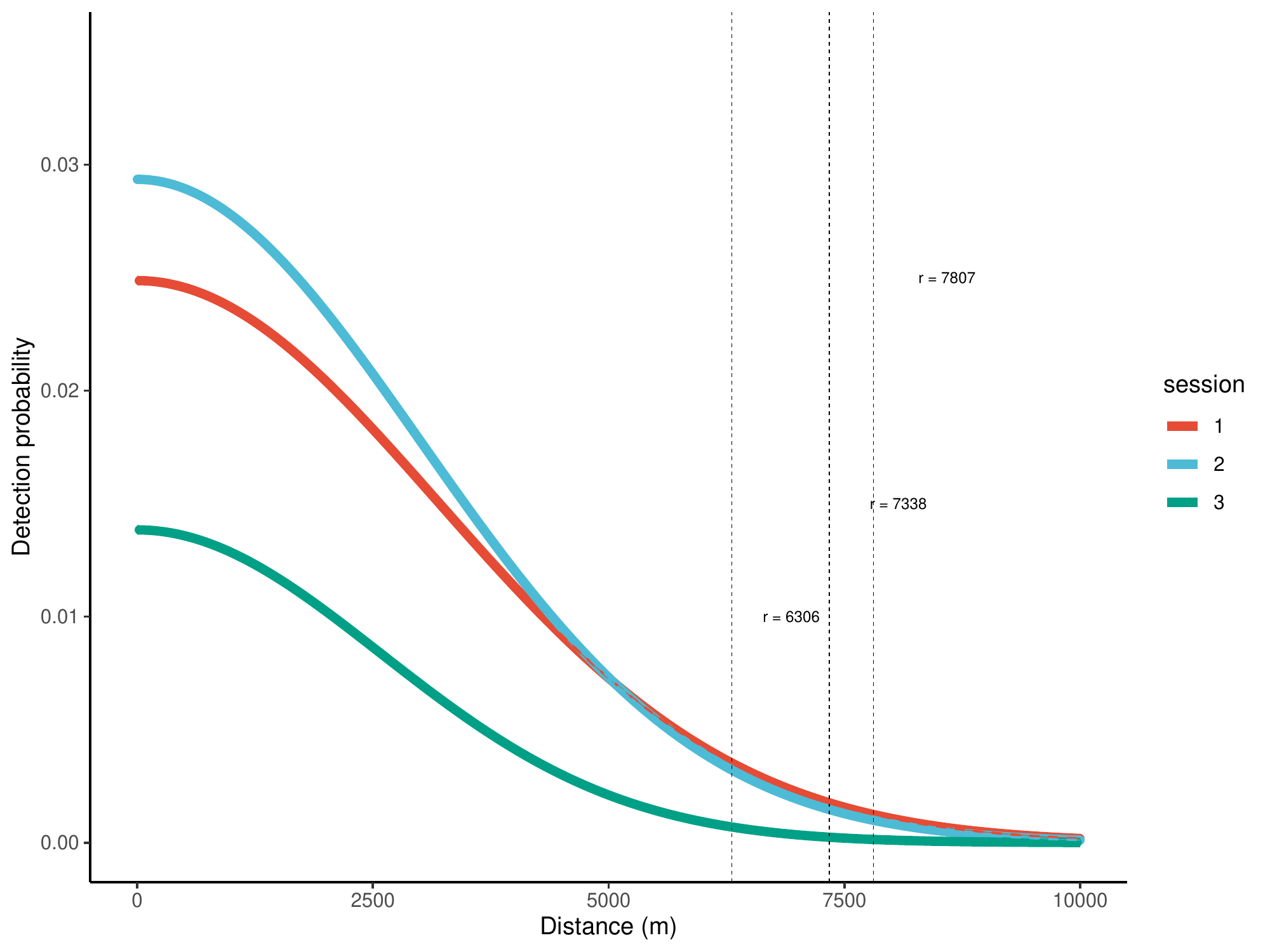}}
  \hfill
  \subfigure[]{\includegraphics{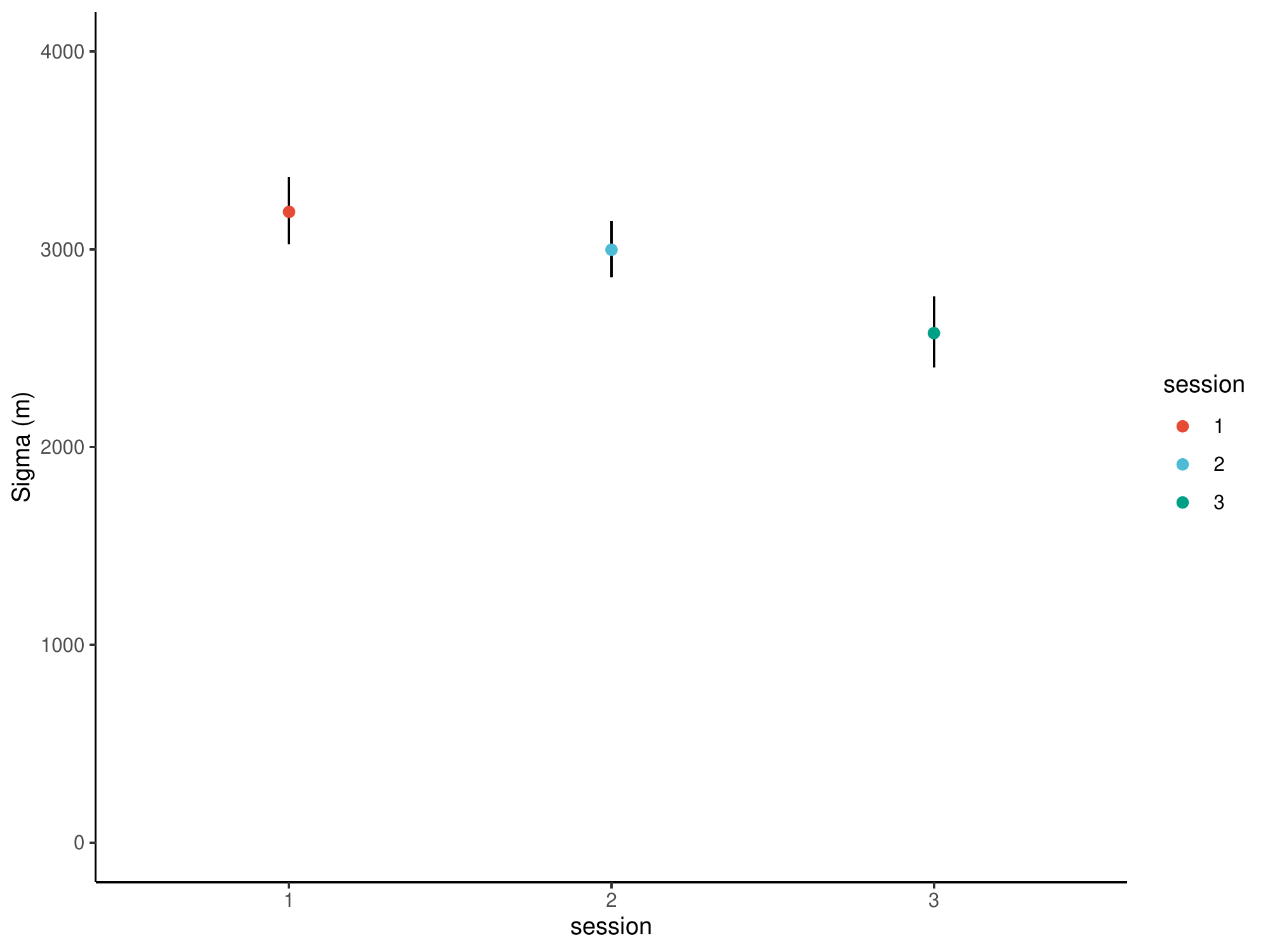}}
  \hfill
  \subfigure[]{\includegraphics{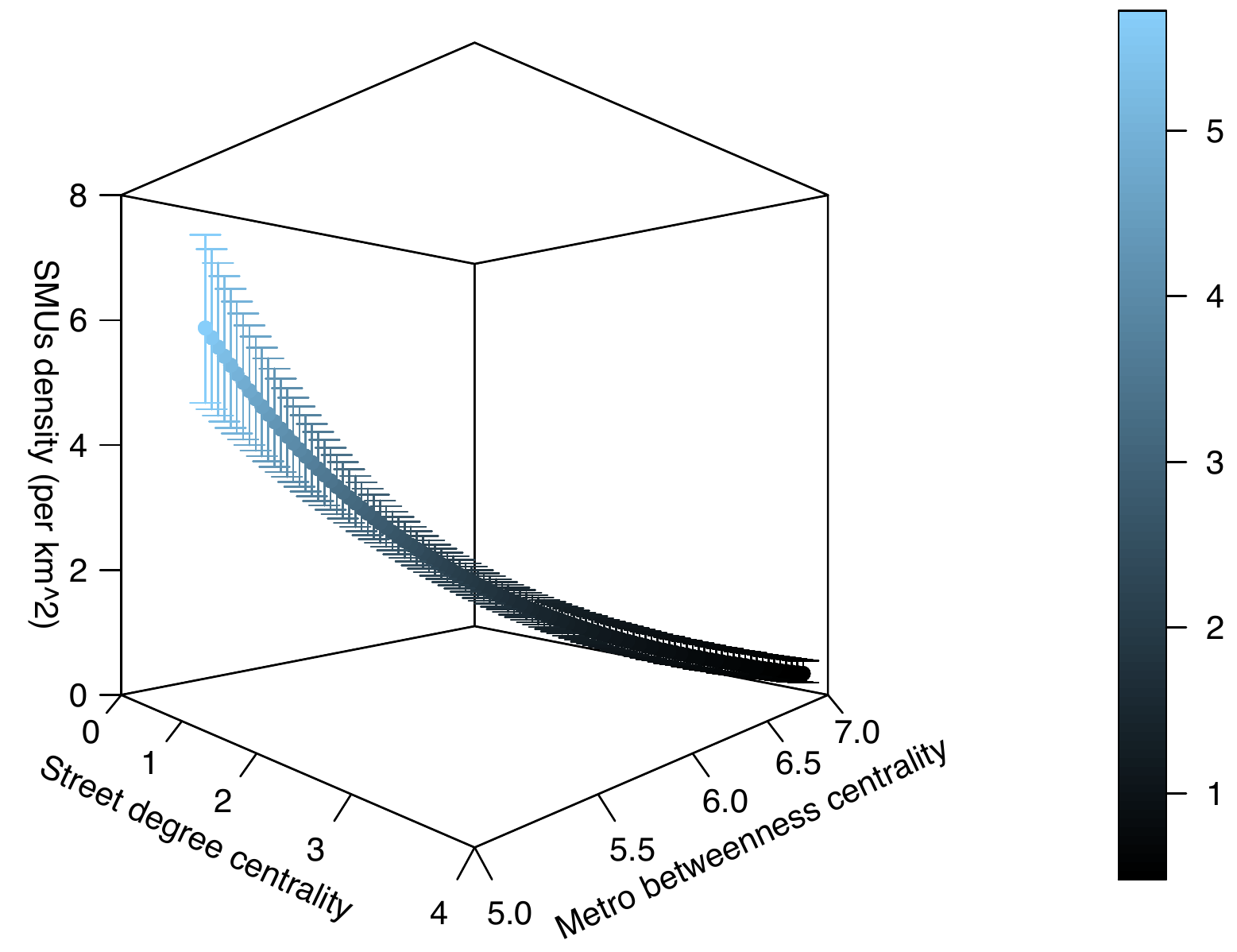}}
  \hfill
\caption{Spatial variation in the detectability and density of SMU protest supporters in Mexico City: \textbf{a}, Baseline detection probability ($p_{0}$). \textbf{b}, Probability of detection ($p$($x$ ,$s$)) and use of space. \textbf{c}, Spatial decay ($\sigma$).  \textbf{d}, Variation in SMU's density ($d(s)$). In Figures \textbf{a}, \textbf{c} and \textbf{d}, the error bar shows the lower and upper prediction intervals}
\label{Variacion}
\end{figure}

The best model found evidence that the baseline detection probability ($p_{0}$) varied in accordance with the \texttt{session}  (i.e., observation day) and time-varying \texttt{Tweetogram}  covariates. The latter case is shown in Figure  ~\ref{Variacion}(a). In this figure, it is possible to observe that the baseline detection probability of  SMUs is higher, at different \texttt{Tweetogram} values, during the day of the protest (i.e. \texttt{session 2}) in comparison to the day before or after the protest event. It is observed that the baseline detectability is higher during the day of the protest event compared to that of a weekday (i.e. \texttt{session 1}). During Sunday (i.e. \texttt{session 3}) there was a noticeable decrease in baseline detectability, as may be observed in the figure. The model thus demonstrates that baseline detection probability varies with the daily post rate in Mexico City and the observation day. Additionally, the model found evidence that $\sigma$ varied with \texttt{session} (see Figure  ~\ref{Variacion}(c)). 

The half-normal function, based on the estimates obtained for $p_{0}$ and $\sigma$,  shows that the detectability (i.e. detection probability) over space of SMU protest supporters is higher during the day of the protest event. Figure  ~\ref{Variacion}(b) describes how the detection probability decays with the distance to the center of activity of the SMUs.  As the figure indicates, the detection probability decays at different rates. The model predicts that the highest probability of detection of SMU protest supporters in Mexico City does not occur at the beginning of the event, nor at the end of the event, but rather at  3:00 p.m., when the \texttt{Tweetogram} obtained a value $\approx$ 0.116 (i.e. when this measure reached its maximum during this day, see Figure  ~\ref{8por16}).

Additionally, the model estimates that the SMUs had a different use of space in the city during the three days of observation. Based on the $\sigma$ value, the 95\% home range radius was calculated as $r_{0.95}$ $=$ $\sigma*\sqrt{5.99}$ (where 5.99 is the chi-square critical value for the 95\% significance level) and the corresponding area used was calculated as $A_{0,95}$ $=$ $\pi$ $r^2$  \citep{Sutherland_2016,Mu_oz_2016}. The day before the protest, $\sigma$ was found to be at its maximum, suggesting that the radius from the SMU's respective activity centres was the largest of the three days ($\sigma$ $=$  3189.74, SE $=$ 86.74; $r_{0.95}$  $=$ 7806.74, $A_{0.95}$ $=$ 191465055). On the day of the protest, the radius was marginally smaller than the previous day ($\sigma$ $=$  2998.22, SE $=$ 73.45; $r_{0.95}$  $=$ 7337.99, $A_{0.95}$ $=$ 169162767) while on the day after the protest, $\sigma$ and therefore the radius was smaller still ($\sigma$ $=$  2576.51, SE $=$ 91.12; $r_{0.95}$  $=$ 6305.89, $A_{0.95}$ $=$ 124923079), what is expected to be observed during a  Sunday (i.e. during a non-working day). These results show that there is variability in the use of space by SMUs and that the use of space is different compared to the day before or after the protest event.

The comparative analysis also uncovered evidence that the spatial density of SMU protest supporters varied with metro and street centralities, as is shown in Figure  ~\ref{Variacion}(d). In the case of street centrality there was an inverse relationship, implying that the lower is $\alpha_{0}\omega$-\texttt{weighted street degree} centrality, the higher is the SMU's density per km$^2$. An inverse relationship between user density and the $\alpha_{0}\omega$-\texttt{weighted metro betweenness} centrality was also found. To obtain a more conservative representation of the functional relationship, we used a 10\% quantile (i.e., $\alpha_{0}\omega$-\texttt{weighted street degree}   $=$ 1.02; $\alpha_{0}\omega$-\texttt{weighted metro betweenness} centrality  $=$ 5.09 ) and a 90\% quantile (i.e., $\alpha_{0}\omega$-\texttt{weighted street degree}   $=$ 3.98; $\alpha_{0}\omega$-\texttt{metro betweenness} centrality $=$ 6.79) to define the prediction range. This means that the centrality of the transportation networks provides information and reference points from which the users' activity centers can be found over space. 

With these settings, the maximum estimated SMU's density was found to be 5.72 SMUs per km$^2$ (SE $=$ 0.62,  lwr $=$ 4.62, upr $=$ 7.09), reached when $\alpha_{0}\omega$-\texttt{weighted street degree}   $=$  1.02, and $\alpha_{0}\omega$-\texttt{weighted metro betweenness} centrality  $=$ 5.08. The minimum estimated SMU's density was found to be 0.47 SMUs per km$^2$ (SE $=$ 0.08,  lwr $=$ 0.34, upr $=$ 0.67), reached when  $\alpha_{0}\omega$-\texttt{weighted street degree}   $=$  3.98, and $\alpha_{0}\omega$-\texttt{weighted metro betweenness} centrality  $=$ 6.79. These results show that the density of users supporting the protest has a dependency on the existing structures of the urban environment of Mexico City. 

\section{Discussion} 

The purpose of this study was to investigate spatial variability in the detectability and density of SMUs in the context of a planned protest event. The strategy consisted of comparing multiple models to find those that best explain the case under study. Our set of hypotheses tested whether variables representing street networks, neighborhood socio-demographic characteristics, social media post rates, and the observation day could constitute a model that fit the observational samples better than the null model or models based on physical proximity to the protest location. In what follows we discuss the main results and findings as they compare with those of the related literature.

One of the most interesting aspects of the results was to find that the detectability of SMUs is highly variable, and how we tested, it does not depend on the physical proximity to the protest location. The results pointed that the baseline detection probability ($p_{0}$) varied with the Twitter post rate as measured by the \texttt{Tweetogram}  metric. The inclusion of this variable greatly improved the detection models compared to the null model. By measuring the daily post rhythm we were able to identify when SMUs were more detectable during the protest event (at 3 p.m.).   We interpret that because the protest march requires the participation of individuals in a certain place and time (i.e., a target-driven user behavior), it results in reduced randomness of movement and posting behavior over space, leading to an increased baseline probability of being detected.   To our knowledge, this empirical finding is one of the first to explain variations in SMU's detectability across space and time in terms of the users' collective posting rhythms. In a previous study \citep{leypunskiy2018geographically}, the \texttt{Tweetogram} measure was used to show mainly that there is a geographical and seasonal variation of user's posting behavior, and in our study, we have demonstrated that using our research design, this measure can be used to predict, at a higher resolution level, where and when users can be observed during the protest event in the city.

Also found by our study was evidence that the spatial decay parameter ($\sigma$) varied with observation day. Of particular interest was how SMU protest supporters occupied space. On the day of the march, users made significant use of the space. This result sheds considerable light on some key aspects of SMU's spatial behavior during the protest event. While the sigma value was significantly different for the day of the protest, we can see that its value was qualitatively similar to that obtained for the day before, a weekday.  On the other hand, the day after the event, SMU's use of space did decline, as is typical on a Sunday when people tend to stay close to home. Our approach thus suggests, based on these empirical findings, that users' activity centre and activity radius favored their participation in the protest event. In other words, users made no extra "effort" to extend their radius of action compared to a weekday.  This can be interpreted to mean that the success of the protest event was due to its location in the center of the city, which is an area where users can reach.

Evidence was found on variations in the spatial density ($d(s)$) of SMUs who supported the protest march analyzed in our case study. The results suggest that the observed densities varied with the street network (i.e., street intersection) degree centrality and metro betweenness centrality. As regards the former measure, the model showed that there were lower densities in areas where centrality was high. This makes sense given that streets in such areas will often have a high confluence of vehicles, pedestrians, or both. In ecological terms, they are not likely places for individuals to establish their activity centre. Therefore, areas with a high street degree centrality and with a high degree of metro betweenness correspond to places where individuals commonly transit.  These network structures can be interpreted as reference points from which the centers of activity of SMUs can be found.

The comparative analysis also showed that the SMU's spatial density did not vary with \texttt{session}. Given that our study covered the entire area of Mexico City and for a short observation period, it is reasonable to expect that the number of SMUs who supported the protest was rather stable. This in turn implies that during the observed period there was no sudden process of generation or reduction of the number of social media supporters to the protest event.  If an increase or a decrease in their density had been observed over the three sessions, a hypothesis on the generation of new activists could have been formulated, but observing such change in a short period was unlikely. At best, an observed increase in density could be interpreted to mean that the individuals turning to social media during the event were already politically active in the offline environment. This is feasible reasoning since a recent repeated-wave panel study found that those individuals who are already politically active are motivated to use social media for participating in political life \citep{oser2020reinforcement}. In other words, the evidence provided here and in previous research suggests that individuals who engage in political activities become active users of social media, not the opposite. In this sense, the best model obtained is showing that there is high variability in the detection of social media users who support a protest, but that there is no major change in density at the city level.\footnote{A clear example where the centres of activity of social media users can change for a short time is the Occupy Wall Street (OWS) protest movement, where hundreds of activists camped out for weeks in Zuccotti Park, New York. An extreme example of how SMU's density can grow rapidly in an area under study is the case of the migrant caravans (i.e. protesting asylum--seekers) along the U.S.-Mexico border. In that type of collective action, density changes can be attributed to the (temporary) changes in the activity centres of highly engaged individuals.} 

The results regarding spatial elements and other user-related factors could, we believe, make a significant theoretical and methodological contribution to human behavior studies. Previous research on protests has stressed the importance of understanding the behavior of SMU protest supporters in terms of their distance from the protest location \citep{wallace2014spatial,cortina2020distance}. Also, previous work has shown that the \textit{local} (i.e. at the protest meeting location) number of individuals participating in a protest can be interpreted as an indicator of the event's magnitude \citep{biggs2018size,sobolevnews,opp2009theories}. For example, some studies on social media have adopted a quasi-experimental approach and difference-in-differences analysis for comparing a group of users exposed to a protest event with a group not so exposed based on their physical distance to the event \citep{Zhang_2016,zhang2016geolocated,karduni2020anatomy}. Other investigations assume that physical distance is an impedance to participation in a protest \citep{Traag_2017,biggs2018size}. In those studies, they do not consider that in the geographical space there are structures of the urban environment that can affect and restrict the spatial behavior of individuals. 

In contrast, our conceptual and methodological approach, inspired by concepts of ecology, chronobiology, network science, and social media research, has pointed up the importance of the spatial dimension in SMU's behavior. The results reveal a completely different picture in which the urban environment and the behavior of social media users have a rich interplay. In the present investigation,  we have focussed on the users themselves, or more specifically, their activity centre as the fundamental factor in modeling their detectability and density.  Our results show that variations in these two phenomena depend on structural elements of the city as well as SMU's Twitter post rhythms. The present study suggests that density is a key factor in that it serves as a global indicator of SMU participants' collective behavior at the city level. But our most theoretically interesting finding is that physical proximity to the protest event location did not prove to be an explanatory variable. This means that generating a theory of physical exposure to the location of the event to explain the behavior of users during the day of the protest event has no empirical support.  Therefore, this research offers a comprehensive analytical framework for generating new insights into spatial user behavior in the context of social media research.

Thinking reflectively, and from a conservative approach, our search for empirical evidence also suggests that there is no general model that can be applied to all cases of protest events, as there is variability in the behavior of individuals and also in the environment they live in.  We believe that future research may consider other potentially relevant factors (e.g. such as gender and age structure of the population, as well as other spatial and geographic covariates or detection functions) would certainly contribute to a better understanding of the sources of variation in the spatial behavior of SMUs. The approach also can help to characterize the factors that delimit spatial behavior of SMUs while protecting the geoprivacy of the user, an increasing requirement for conducting research based on geotagged social media data as discussed by \citet{hu2020understanding}. 

To sum up, the evidence-based comparative models generated by our study, have identified the patterns of variation in spatial detectability and density determined by transport networks and the users' own internal cognitive mechanisms as represented here by the \texttt{Tweetogram} measure. It was found that the rate of daily social media posts and the observation day are more powerful in detecting social media users than street network centralities or measures based on relative wealth or population density indexes. On the day of the protest, users exhibited maximum detectability and significant use of space. Evidence was also found that the density of social media users who supported the protest can be explained by metro and street network centrality measures. It was further demonstrated that information relating to physical proximity to the protest location, taken by itself, performed worse than the null model in modeling the spatial detectability and density of SMUs. 

Overall, our study provides consistent evidence that the information embedded in the geographic space where users post social media content does matter. The built environment and the collective posting rate are significant explanatory factors of the variations in their spatial detectability and density of social media user protest supporters. Changes observed in these two elements may be interpreted as indicators of users' level of engagement during protest events. Finally, we encourage other researchers, especially those focused on the problem of ecological validity, to explore and propose new hypotheses on the rich and complex behavior of individuals in their online and offline environments.

\section*{Funding}

This work was supported by the Deutsche Forschungsgemeinschaft (DFG) under grant \textsc{GRK 2167}, Research Training Group ``\textsc{User-Centred Social Media}" (\textsc{UCSM}).

\section*{Declaration of conflicting interests}

The author(s) declared no potential conflicts of interest concerning the research, authorship, and/or publication of this article.

\section*{Acknowledgements}

We acknowledge the support of the \textsc{National Laboratory of High-Performance Computing} (NLHPC),  hosted at the \textsc{Center for Mathematical Modeling} (CMM, Universidad de Chile).

\bibliographystyle{unsrtnat}
\bibliography{PAPER.bib}  

\begin{thebibliography}{49}
\providecommand{\natexlab}[1]{#1}
\providecommand{\url}[1]{\texttt{#1}}
\expandafter\ifx\csname urlstyle\endcsname\relax
  \providecommand{\doi}[1]{doi: #1}\else
  \providecommand{\doi}{doi: \begingroup \urlstyle{rm}\Url}\fi

\bibitem[Fisher et~al.(2019)Fisher, Andrews, Caren, Chenoweth, Heaney, Leung,
  Perkins, and Pressman]{fisher2019science}
Dana~R. Fisher, Kenneth~T. Andrews, Neal Caren, Erica Chenoweth, Michael~T.
  Heaney, Tommy Leung, L.~Nathan Perkins, and Jeremy Pressman.
\newblock The science of contemporary street protest: New efforts in the
  {United States}.
\newblock \emph{Science Advances}, 5\penalty0 (10):\penalty0 eaaw5461, oct
  2019.

\bibitem[Mooijman et~al.(2018)Mooijman, Hoover, Lin, Ji, and
  Dehghani]{mooijman2018moralization}
Marlon Mooijman, Joe Hoover, Ying Lin, Heng Ji, and Morteza Dehghani.
\newblock Moralization in social networks and the emergence of violence during
  protests.
\newblock \emph{Nature Human Behaviour}, 2\penalty0 (6):\penalty0 389--396, may
  2018.

\bibitem[Chen and Pirolli(2012)]{chen2012you}
Jilin Chen and Peter Pirolli.
\newblock Why you are more engaged: {factors} influencing twitter engagement in
  {Occupy Wall Street}.
\newblock In \emph{Proceedings of the Sixth International AAAI Conference on
  Weblogs and Social Media}, volume~6, pages 423--426. AAAI Press, 2012.

\bibitem[Traag et~al.(2017)Traag, Quax, and Sloot]{Traag_2017}
V.A. Traag, R.~Quax, and P.M.A. Sloot.
\newblock Modelling the distance impedance of protest attendance.
\newblock \emph{Physica A: Statistical Mechanics and its Applications},
  468:\penalty0 171--182, feb 2017.

\bibitem[Barber{\'{a}} et~al.(2015)Barber{\'{a}}, Wang, Bonneau, Jost, Nagler,
  Tucker, and Gonz{\'{a}}lez-Bail{\'{o}}n]{Barber__2015}
Pablo Barber{\'{a}}, Ning Wang, Richard Bonneau, John~T. Jost, Jonathan Nagler,
  Joshua Tucker, and Sandra Gonz{\'{a}}lez-Bail{\'{o}}n.
\newblock The critical periphery in the growth of social protests.
\newblock \emph{{PLOS} {ONE}}, 10\penalty0 (11):\penalty0 e0143611, nov 2015.

\bibitem[Davies et~al.(2013)Davies, Fry, Wilson, and
  Bishop]{davies2013mathematical}
Toby~P. Davies, Hannah~M. Fry, Alan~G. Wilson, and Steven~R. Bishop.
\newblock A mathematical model of the {L}ondon riots and their policing.
\newblock \emph{Scientific Reports}, 3\penalty0 (1):\penalty0 1--9, feb 2013.

\bibitem[Lemos et~al.(2016)Lemos, Coelho, and Lopes]{lemos2017protestlab}
Carlos~M. Lemos, Helder Coelho, and Rui~J. Lopes.
\newblock {ProtestLab}: A computational laboratory for studying street
  protests.
\newblock In \emph{Nonlinear Systems and Complexity}, pages 3--29. Springer
  International Publishing, dec 2016.

\bibitem[Pires and Crooks(2017)]{pires2017modeling}
Bianica Pires and Andrew~T. Crooks.
\newblock Modeling the emergence of riots: A geosimulation approach.
\newblock \emph{Computers, Environment and Urban Systems}, 61:\penalty0 66--80,
  jan 2017.

\bibitem[Bacaksizlar(2019)]{bacaksizlar2019understanding}
Nazmiye~Gizem Bacaksizlar.
\newblock \emph{Understanding Social Movements through Simulations of Anger
  Contagion in Social Media}.
\newblock PhD thesis, The University of North Carolina at Charlotte, 2019.

\bibitem[McKinney(2008)]{mckinney2008effects}
Michael~L. McKinney.
\newblock Effects of urbanization on species richness: A review of plants and
  animals.
\newblock \emph{Urban Ecosystems}, 11\penalty0 (2):\penalty0 161--176, jan
  2008.

\bibitem[Sahu et~al.(2019)Sahu, Parganiha, and Pati]{sahu2019spatiotemporal}
Bhupendra~Kumar Sahu, Arti Parganiha, and Atanu~Kumar Pati.
\newblock Spatiotemporal variability in activity patterns of urban street
  cattle as function of environmental factors.
\newblock \emph{Chronobiology International}, 36\penalty0 (10):\penalty0
  1362--1372, aug 2019.

\bibitem[Arnaiz-Schmitz et~al.(2018)Arnaiz-Schmitz, Schmitz,
  Herrero-J{\'{a}}uregui, Guti{\'{e}}rrez-Angonese, Pineda, and
  Montes]{arnaiz2018identifying}
C.~Arnaiz-Schmitz, M.F. Schmitz, C.~Herrero-J{\'{a}}uregui,
  J.~Guti{\'{e}}rrez-Angonese, F.D. Pineda, and C.~Montes.
\newblock Identifying socio-ecological networks in rural-urban gradients:
  Diagnosis of a changing cultural landscape.
\newblock \emph{Science of The Total Environment}, 612:\penalty0 625--635, jan
  2018.

\bibitem[Valdez(2018)]{valdez2019homeostatic}
Pablo Valdez.
\newblock Homeostatic and circadian regulation of cognitive performance.
\newblock \emph{Biological Rhythm Research}, 50\penalty0 (1):\penalty0 85--93,
  jul 2018.

\bibitem[Fischer et~al.(2019)Fischer, Stefano, Gras, Kramer-Schadt, Sutherland,
  Coulson, and Stillfried]{fischer2019circadian}
Manuela Fischer, Julian~Di Stefano, Pierre Gras, Stephanie Kramer-Schadt,
  Duncan~R. Sutherland, Graeme Coulson, and Milena Stillfried.
\newblock Circadian rhythms enable efficient resource selection in a
  human-modified landscape.
\newblock \emph{Ecology and Evolution}, 9\penalty0 (13):\penalty0 7509--7527,
  jun 2019.

\bibitem[Leypunskiy et~al.(2018)Leypunskiy, K{\i}c{\i}man, Shah, Walch,
  Rzhetsky, Dinner, and Rust]{leypunskiy2018geographically}
Eugene Leypunskiy, Emre K{\i}c{\i}man, Mili Shah, Olivia~J. Walch, Andrey
  Rzhetsky, Aaron~R. Dinner, and Michael~J. Rust.
\newblock Geographically resolved rhythms in {Twitter} use reveal social
  pressures on daily activity patterns.
\newblock \emph{Current Biology}, 28\penalty0 (23):\penalty0 3763--3775.e5, dec
  2018.

\bibitem[Diwan et~al.(2020)Diwan, Swain, Minz, Parganiha, and Pati]{Diwan_2020}
Ananya Diwan, Rakesh~Kumar Swain, Sarojini Minz, Arti Parganiha, and
  Atanu~Kumar Pati.
\newblock Ultradian, circadian, and circaseptan rhythms in the patterns of
  usage of {Facebook} messenger.
\newblock \emph{Biological Rhythm Research}, pages 1--9, apr 2020.

\bibitem[Murnane et~al.(2015)Murnane, Abdullah, Matthews, Choudhury, and
  Gay]{murnane2015social}
Elizabeth~L Murnane, Saeed Abdullah, Mark Matthews, Tanzeem Choudhury, and Geri
  Gay.
\newblock Social (media) jet lag: How usage of social technology can modulate
  and reflect circadian rhythms.
\newblock In \emph{Proceedings of the 2015 {ACM} International Joint Conference
  on Pervasive and Ubiquitous Computing - {UbiComp} {\textquotesingle}15},
  pages 843--854. {ACM} Press, 2015.

\bibitem[Swain and Pati(2019)]{kumar2020habitual}
Rakesh~Kumar Swain and Atanu~Kumar Pati.
\newblock Habitual daily `{Good Morning}' message senders reveal the status of
  their own circadian clock.
\newblock \emph{Biological Rhythm Research}, pages 1--12, jan 2019.

\bibitem[Rui and Ban(2014)]{Rui_2014}
Yikang Rui and Yifang Ban.
\newblock Exploring the relationship between street centrality and land use in
  {S}tockholm.
\newblock \emph{International Journal of Geographical Information Science},
  28\penalty0 (7):\penalty0 1425--1438, mar 2014.

\bibitem[Bielik et~al.(2018)Bielik, König, Schneider, and
  Varoudis]{Bielik_2018}
M.~Bielik, R.~König, S.~Schneider, and T.~Varoudis.
\newblock Measuring the impact of street network configuration on the
  accessibility to people and walking attractors.
\newblock \emph{Networks and Spatial Economics}, 18\penalty0 (3):\penalty0
  657--676, sep 2018.

\bibitem[Summers and Johnson(2016)]{Summers_2016}
Lucia Summers and Shane~D. Johnson.
\newblock Does the configuration of the street network influence where outdoor
  serious violence takes place? {U}sing space syntax to test crime pattern
  theory.
\newblock \emph{Journal of Quantitative Criminology}, 33\penalty0 (2):\penalty0
  397--420, jun 2016.

\bibitem[Nieves et~al.(2017)Nieves, Stevens, Gaughan, Linard, Sorichetta,
  Hornby, Patel, and Tatem]{Nieves_2017}
Jeremiah~J. Nieves, Forrest~R. Stevens, Andrea~E. Gaughan, Catherine Linard,
  Alessandro Sorichetta, Graeme Hornby, Nirav~N. Patel, and Andrew~J. Tatem.
\newblock Examining the correlates and drivers of human population
  distributions across low- and middle-income countries.
\newblock \emph{Journal of The Royal Society Interface}, 14\penalty0
  (137):\penalty0 20170401, dec 2017.

\bibitem[Nadai et~al.(2016)Nadai, Staiano, Larcher, Sebe, Quercia, and
  Lepri]{De_Nadai_2016}
Marco~De Nadai, Jacopo Staiano, Roberto Larcher, Nicu Sebe, Daniele Quercia,
  and Bruno Lepri.
\newblock The death and life of great italian cities.
\newblock In \emph{Proceedings of the 25th International Conference on World
  Wide Web - {WWW} {\textquotesingle}16}, pages 413--423. {ACM} Press, 2016.

\bibitem[Liu et~al.(2020)Liu, Fang, and Jing]{Liu_2020}
Yaolin Liu, Feiguo Fang, and Ying Jing.
\newblock How urban land use influences commuting flows in {Wuhan}, {Central
  China}: A mobile phone signaling data perspective.
\newblock \emph{Sustainable Cities and Society}, 53:\penalty0 101914, feb 2020.

\bibitem[Royle et~al.(2013)Royle, Chandler, Sollmann, and
  Gardner]{royle2013spatial}
J~Andrew Royle, Richard~B Chandler, Rahel Sollmann, and Beth Gardner.
\newblock \emph{Spatial capture-recapture}.
\newblock Academic Press, 2013.

\bibitem[Bosia et~al.(2019)Bosia, McEvoy, and Rahman]{bosia2020oxford}
Michael~J. Bosia, Sandra~M. McEvoy, and Momin Rahman, editors.
\newblock \emph{The Oxford Handbook of Global {LGBT} and Sexual Diversity
  Politics}.
\newblock Oxford University Press, jan 2019.

\bibitem[Beer and Cruz-Aceves(2018)]{beer2018extending}
Caroline Beer and Victor~D. Cruz-Aceves.
\newblock Extending rights to marginalized minorities: Same-sex relationship
  recognition in {Mexico} and the {United States}.
\newblock \emph{State Politics {\&} Policy Quarterly}, 18\penalty0
  (1):\penalty0 3--26, jan 2018.

\bibitem[Walker(2014)]{walker2014equality}
Cameron Walker.
\newblock Equality: Standing out.
\newblock \emph{Nature}, 505\penalty0 (7482):\penalty0 249--251, jan 2014.

\bibitem[Boeing(2017)]{boeing2017osmnx}
Geoff Boeing.
\newblock {OSMnx}: New methods for acquiring, constructing, analyzing, and
  visualizing complex street networks.
\newblock \emph{Computers, Environment and Urban Systems}, 65:\penalty0
  126--139, sep 2017.

\bibitem[Boeing(2020)]{Boeing_2020}
Geoff Boeing.
\newblock The right tools for the job: The case for spatial science
  tool-building.
\newblock \emph{Transactions in {GIS}}, 24\penalty0 (5):\penalty0 1299--1314,
  sep 2020.

\bibitem[Opsahl et~al.(2010)Opsahl, Agneessens, and Skvoretz]{opsahl2010node}
Tore Opsahl, Filip Agneessens, and John Skvoretz.
\newblock Node centrality in weighted networks: Generalizing degree and
  shortest paths.
\newblock \emph{Social Networks}, 32\penalty0 (3):\penalty0 245--251, jul 2010.

\bibitem[Royle et~al.(2017)Royle, Fuller, and Sutherland]{royle2018unifying}
J.~Andrew Royle, Angela~K. Fuller, and Christopher Sutherland.
\newblock Unifying population and landscape ecology with spatial
  capture-recapture.
\newblock \emph{Ecography}, 41\penalty0 (3):\penalty0 444--456, aug 2017.

\bibitem[Rotman and Shalev(2020)]{rotman2020using}
Assaf Rotman and Michael Shalev.
\newblock Using location data from mobile phones to study participation in mass
  protests.
\newblock \emph{Sociological Methods {\&} Research}, page 004912412091492, apr
  2020.

\bibitem[Sutherland et~al.(2019)Sutherland, Royle, and
  Linden]{sutherland2019oscr}
Chris Sutherland, J.~Andrew Royle, and Daniel~W. Linden.
\newblock {oSCR}: {a} spatial capture{\textendash}recapture {R} package for
  inference about spatial ecological processes.
\newblock \emph{Ecography}, 42\penalty0 (9):\penalty0 1459--1469, jul 2019.

\bibitem[Masias et~al.(2019)Masias, Hecking, Crespo, and Hoppe]{Masias_2019}
Victor~H. Masias, Tobias Hecking, Fernando Crespo, and H.~Ulrich Hoppe.
\newblock Detecting social media users based on pedestrian networks and
  neighborhood attributes: an observational study.
\newblock \emph{Applied Network Science}, 4\penalty0 (1), oct 2019.

\bibitem[Santos et~al.(2018)Santos, Soares, Abreu, Araujo, and
  Santos]{santos2018cross}
Miriam~Seoane Santos, Jastin~Pompeu Soares, Pedro~Henrigues Abreu, Helder
  Araujo, and Joao Santos.
\newblock Cross-validation for imbalanced datasets: Avoiding overoptimistic and
  overfitting approaches [research frontier].
\newblock \emph{{IEEE} Computational Intelligence Magazine}, 13\penalty0
  (4):\penalty0 59--76, nov 2018.

\bibitem[Mas{\'{\i}}as et~al.(2016)Mas{\'{\i}}as, Baldwin, Laengle, Vargas, and
  Crespo]{masias2017exploring}
V{\'{\i}}ctor~Hugo Mas{\'{\i}}as, Paula Baldwin, Sigifredo Laengle, Augusto
  Vargas, and Fernando~A. Crespo.
\newblock Exploring the prominence of {R}omeo and {J}uliet's characters using
  weighted centrality measures.
\newblock \emph{Digital Scholarship in the Humanities}, page fqw029, aug 2016.

\bibitem[Sutherland et~al.(2016)Sutherland, Mu{\~{n}}oz, Miller, and
  Grant]{Sutherland_2016}
Chris Sutherland, David~J. Mu{\~{n}}oz, David~A.W. Miller, and Evan H.~Campbell
  Grant.
\newblock Spatial capture{\textendash}recapture: A promising method for
  analyzing data collected using artificial cover objects.
\newblock \emph{Herpetologica}, 72\penalty0 (1):\penalty0 6--12, mar 2016.

\bibitem[Mu{\~{n}}oz et~al.(2016)Mu{\~{n}}oz, Miller, Sutherland, and
  Grant]{Mu_oz_2016}
David~J. Mu{\~{n}}oz, David A.~W. Miller, Chris Sutherland, and Evan
  H.~Campbell Grant.
\newblock Using spatial capture{\textendash}recapture to elucidate population
  processes and space-use in herpetological studies.
\newblock \emph{Journal of Herpetology}, 50\penalty0 (4):\penalty0 570--581,
  dec 2016.

\bibitem[Oser and Boulianne(2020)]{oser2020reinforcement}
Jennifer Oser and Shelley Boulianne.
\newblock Reinforcement effects between digital media use and political
  participation: A meta-analysis of repeated-wave panel data.
\newblock \emph{Public Opinion Quarterly}, 84\penalty0 (S1):\penalty0 355--365,
  2020.

\bibitem[Wallace et~al.(2014)Wallace, Zepeda-Mill{\'a}n, and
  Jones-Correa]{wallace2014spatial}
Sophia~J Wallace, Chris Zepeda-Mill{\'a}n, and Michael Jones-Correa.
\newblock Spatial and temporal proximity: Examining the effects of protests on
  political attitudes.
\newblock \emph{American Journal of Political Science}, 58\penalty0
  (2):\penalty0 433--448, 2014.

\bibitem[Cortina(2019)]{cortina2020distance}
Jeronimo Cortina.
\newblock From a distance: Geographic proximity, partisanship, and public
  attitudes toward the {U.S.{\textendash}Mexico} border wall.
\newblock \emph{Political Research Quarterly}, 73\penalty0 (3):\penalty0
  740--754, jun 2019.

\bibitem[Biggs(2016)]{biggs2018size}
Michael Biggs.
\newblock Size matters.
\newblock \emph{Sociological Methods {\&} Research}, 47\penalty0 (3):\penalty0
  351--383, feb 2016.

\bibitem[Sobolev et~al.(2020)Sobolev, Chen, Joo, Steinert-Threlkeld,
  et~al.]{sobolevnews}
Anton Sobolev, M~Keith Chen, Jungseock Joo, Zachary~C Steinert-Threlkeld,
  et~al.
\newblock News and geolocated social media accurately measure protest size
  variation.
\newblock \emph{American Political Science Review}, 114\penalty0 (4):\penalty0
  1343--1351, jun 2020.

\bibitem[Opp(2009)]{opp2009theories}
Karl-Dieter Opp.
\newblock \emph{Theories of Political Protest and Social Movements}.
\newblock Routledge, apr 2009.

\bibitem[Zhang(2016)]{Zhang_2016}
Han Zhang.
\newblock Physical exposures to political protests impact civic engagement:
  Evidence from 13 quasi-experiments with chinese social media.
\newblock \emph{{SSRN} Electronic Journal}, 2016.

\bibitem[Zhang et~al.(2016)Zhang, Hill, and Rothschild]{zhang2016geolocated}
Han Zhang, Shawndra Hill, and David Rothschild.
\newblock Geolocated twitter panels to study the impact of events.
\newblock In \emph{2016 AAAI Spring Symposium Series}. AAAI press, Palo Alto.,
  2016.

\bibitem[Karduni and Sauda(2020)]{karduni2020anatomy}
Alireza Karduni and Eric Sauda.
\newblock Anatomy of a protest: Spatial information, social media, and urban
  space.
\newblock \emph{Social Media and Society}, 6\penalty0 (1):\penalty0
  205630511989732, jan 2020.

\bibitem[Hu and Wang(2020)]{hu2020understanding}
Yingjie Hu and Ruo-Qian Wang.
\newblock Understanding the removal of precise geotagging in tweets.
\newblock \emph{Nature Human Behaviour}, 4\penalty0 (12):\penalty0 1219--1221,
  sep 2020.

\end{thebibliography}

\end{document}